\documentclass[onecolumn]{aastex62}
\usepackage{amsmath}
\usepackage[switch]{lineno}
\usepackage{diagbox}
\usepackage{bm}

\received{, 2023}
\revised{, 2023}
\accepted{, 2023}

\submitjournal{ApJL}

\shorttitle{SOC in GRB precursors}
\shortauthors{Li et al.}

\begin{document}
\title{Signatures of the Self-organized Criticality Phenomenon in Precursors of Gamma-ray bursts}
\author[0000-0001-6469-8725]{Li, Xiu-Juan$^{\ddag}$}
\affiliation{School of Cyber Science and Engineering, Qufu Normal University, Qufu 273165, China}
\affiliation{Key Laboratory of Modern Astronomy and Astrophysics (Nanjing University), Ministry of Education, Nanjing 210023, P. R. China}
\author[0000-0001-6469-8725]{Yang, Yu-Peng}
\affiliation{School of Physics and Physical Engineering, Qufu Normal University, Qufu 273165, China}
\email{lxj@qfnu.edu.cn}
\begin{abstract}
Precursors provide important clues to the nature of gamma-ray burst (GRB) central engines and can be used to contain GRB physical processes. In this letter, we study the self-organized criticality in precursors of long GRBs in the third Swift/BAT Catalog. We investigate the differential and cumulative size distributions of 100 precursors, including peak flux, duration, rise time, decay time, and quiescent time with the Markov Chain Monte Carlo technique. It is found that all of the distributions can be well described by power-law models and understood within the physical framework of a self-organized criticality system. In addition, we inspect the cumulative distribution functions of the
size differences with a $q$-Gaussian function. The scale-invariance structures of precursors further strengthen our findings. Particularly, similar analyses are made in 127 main bursts. The results show that both precursors and main bursts can be attributed to an self-organized criticality system with the spatial dimension $S = 3$ and driven by the similar magnetically dominated process.
\end{abstract}
\keywords{
\href{http://astrothesaurus.org/uat/739}{High energy astrophysics (739)};
\href{http://astrothesaurus.org/uat/629}{Gamma-ray bursts (629)};
}

\section{Introduction} \label{sec:intro}
In a fraction of gamma-ray bursts (GRBs), some weaker pulses prior to main bursts have been detected, which are usually called as precursors.
Single and multiple precursors have been found in both short GRBs and long GRBs \cite[e.g.][]{Troja2010,Hu2014,Li2022}. Some studies reported that there is no obvious difference between precursors and main bursts by the temporal and spectral analysis, indicating that they might originate from the same mechanism and share the same origin \cite[e.g.][]{Burlon2008,Burlon2009,Lan2018,Lxj2021,Lxj2022}. Some theoretical models have been proposed to explain the origin of precursors, involving photospheric precursors \citep [e.g.,][]{Murakami1991,Lyutikov2000,Daigne2002}, shock breakout precursors \citep [e.g.,][]{Ramirez2002,Wang2007} and so on. It is expected that precursors provide important clues as to the nature of the central engines of GRBs, and more importantly, unveil the physical mechanism of themselves, which has so far remained mysterious.

For a self-organized criticality (SOC) system, subsystem with a nonlinear
energy dissipation process in an external field assembles itself into a critical state. Owing to a small perturbation, the critical state is triggered and will generate an avalanche-like chain reaction of any size \citep{Bak1987}. Typical characteristics of an SOC system are that the event sizes present scale-free power-law-like distributions \citep{Aschwanden2011,Wang2013,Yi2016,Zhang2022}.
A universal statistical analytical SOC model developed by \cite{Aschwanden2012} provided an theoretical framework to quantitatively connect the concept of fractal dimensions to the power-law distributions. In addition, another remarkable hallmark of the SOC system is the scale-invariance structure of the avalanche size differences, which can be described by the probability density functions (PDFs) with a invariant $q$-Gaussian form \citep{Caruso2007,Wei2021,Wang2023}.

\cite{Aschwanden2022} reported that the dynamical nonlinear behaviors of almost all of the discipline in astrophysical systems are governed by SOC process.
As the most luminous explosion events in the universe, the physical processes and radiation mechanisms
of GRBs are still debated. The sizes frequency distributions of GRB X-ray flares, including the energies, durations, peak fluxes, rise times, decay times and waiting times, etc., have been reported to be
attributed to a magnetic reconnection process on the basis of the fractal-diffusive avalanche \citep{Wang2013,Yi2016,Wei2023}.
Besides X-ray flares, similar size distributions of GRB optical flares are found by \cite{Yi2017}, indicating the similar SOC framework to X-ray flares. It is generally believed that X-ray flares may be internal
emission through the collisions of relativistic shells ejected due to the delayed central engine activities \citep[e.g.][]{Burrows2005,ZhangB2006,M2006}. Thus, it is necessary to investigate whether the GRB prompt phases can also be attributed to an SOC process. It is worth noting that the SOC behavior in GRB prompt phase have been studied by \cite{Lv2020} and the results tentatively suggested that SOC phenomena may exist. However, only the GRBs with at least three pulses have been investigated yet. In practice, a fraction of short GRBs are overwhelmingly single-peaked
and double-peaked bursts \citep{Hakkila2018,Lxj2020,Lxj2021}, hence we have examined in detail the properties of the single-peaked and double-peaked BATSE and Swift short GRBs and confirmed the SOC behaviors in short GRB prompt phases in our recent work \citep[][hereafter Paper I]{Lxj6}.

As an important component of GRB prompt phase, precursor provides important information on the nature of central engines.
However, the indisputable fact is that no work has been done yet to disclose whether the SOC features also exists in GRB precursors.
Therefore, we perform a analysis to search for the evidences of the SOC behaviors in precursors in this letter. We focus on the size distributions and further check the scale-invariance structure of the size differences. In addition, we also investigate the two kinds of distributions of main bursts to examine the possible connection between them.
This paper is organized as follows.
Data analysis and methods are given in Section 2. Our main results are presented in Section 3, in which we will give
special attention to a direct comparison with the results of Paper I.
Finally, the discussions and conclusions are presented in Sections 4 and 5.

\section{DATA and Method} \label{sec:observations}
Our precursor sample is from \cite{Li2022}. They limited the
bursts to events with durations longer than 2 s from 1006 GRBs in the third Swift/BAT catalog. The signature of the precursor sample should be clearly shown in all Swift energy-channel 64 ms binning lightcurves, including 15 - 25, 25 - 50, 50 - 100, 100 - 350, and 15 - 350 keV. Finally, they identified 52 long GRBs with the precursor activity. In total, they obtained 227 pulses, including 127 main pulses and 100 precursor pulses. Then, they use the fitting function provided by \cite{Norris2005} to fit all the precursor and the main burst pulses. In addition, peak-time intervals and quiescent times between different episodes were extracted. The detailed sample information and data processing refers to \cite{Li2022}. In this letter, we select the characteristic parameters of the precursor pulses to analyse, including peak flux ($P$), duration ($D$), rise time ($t_ {\rm r}$), and decay time ($t_ {\rm d}$). Besides these parameters, we also investigate 52 peak-time intervals ($\Delta t_ {\rm p1}$) and 52 quiescent times ($\Delta t_ {\rm p2}$) between precursors and main bursts. Note that considering the limited number of GRBs with more than two precursors from \cite{Li2022}, we thus only select the peak-time intervals and quiescent times of GRBs with one precursor.

As applied in our Paper I, an thresholded size distribution function motivated by
physical arguments (e.g. instability threshold, detection threshold, background noise) \citep{Aschwanden2015} will be utilized again, which can be written as
\begin{equation} \label{eq1}
N_{\rm d}(x)=\frac{dN}{dx}\propto(x_{0d}+x)^{-\alpha_{d}}, x_1 \leq x \leq x_2,
\end{equation}
where $N$ is the number of events, $\alpha_{\rm d}$ is the power-law index, $x_{\rm 1}$ represents the minimum cutoff values of scale-free range, $x_{\rm 2}$ is the upper bound by the largest event, which represents a sharp cutoff in the differential size distribution, and $x_{\rm {0d}}$ is a threshold value of differential size distribution, which is common in
astrophysical data sets and represents the deviations from ideal power law size distributions \citep{Newman2005} caused by some natural effects.
$\sigma _{d}=\sqrt{N_{bin,i}}/\triangle x_i$ represents the uncertainty of the differential size distribution, where $N_{bin,i}$ and $\triangle x_i$ are the number of events of the $i$-bin and the bin size, respectively. Note that such the thresholded
power law size distribution is also called a ``Generalized
Pareto distribution'', the ``Generalized Pareto Type II distribution'' and the ``Lomax distribution'' \citep[e.g.][]{Hosking1987, Johnson1994,Lomax1954}. For small data sets and near the upper cutoff, it is statistically advantageous to
use the cumulative size distribution of the Equation \ref {eq1} \citep{Aschwanden2015}, which includes all events accumulated above some
size $x$ and can be obtained as
\begin{equation} \label{eq2}
N_{\rm c}(>x)=1+(N_{ev}-1) \times \frac{(x_2+x_{0c})^{1-\alpha_c}-(x+x_{\rm 0c})^{1-\alpha_c}}{(x_2+x_{0c})^{1-\alpha_c }-(x_1+x_{0c})^{1-\alpha_{c}}},
\end{equation}
where $N_{ev}$ refers to the total number of events, $\alpha_{\rm c}\neq 1$ is the power-law index, and $x_{\rm {0c}}$ is a threshold value of cumulative size distribution. The uncertainty of the cumulative distribution in a given bin $i$ is estimated with $\sigma _{c}=\sqrt{N_{i} }$, where $N_{\rm i}$ is the number of events of the bin. Due to the small sample, a rank-order plot which is essentially an optimum adjustment to minimum statistics is used to get the size distributions. The detailed method refers to
the related descriptions in \cite{Aschwanden2011,Aschwanden2019}.

The Markov chain Monte Carlo (MCMC) technique is first utilized to obtain the best-fit results with the package $pymc$. Then, the standard reduced chi-square ($\chi{_{\rm \nu} ^2}$) criterion which is more suitable for modeling astrophysical data with some natural effects is used to test the goodness of fits \citep{Aschwanden2019}, although it may give biased and noisy slope estimates \citep{Goldstein2004,Bauke2007}. The $\chi{_{\rm \nu} ^2}$ criterions can be written as
\begin{equation} \label{eq3}
\chi{_{\nu,d}^2}=\frac{1}{(n_x-n_{p})} \sum \limits_{i=1}^{n_x} \frac{{[N_{fit,d}(x_i)-N_{obs,d}(x_i)]}^2}{\sigma_{d,i}^2}
\end{equation}
for the differential distribution function, and
\begin{equation} \label{eq3}
\chi{_{\nu,c}^2}=\frac{1}{(n_x-n_{p})} \sum \limits_{i=1}^{n_x} \frac{{[N_{fit,c}(x_i)-N_{obs,c}(x_i)]}^2}{\sigma_{c,i}^2}
\end{equation}
for the cumulative distribution function \citep{Aschwanden2015},
where $n_{\rm x}$ is the number of logarithmic bins, $n_{\rm p}$ is the number of the free parameters, $N_{\rm fit,d}(x_i)$ and $N_{\rm fit,c}(x_i)$ are the theoretical values, and $N_{\rm obs,d}(x_i)$ and $N_{\rm obs,c}(x_i)$ are the corresponding observed values, respectively. Note that the points below the threshold $x_0$ are just noise and do not contribute to the accuracy of the best-fit power-law index, thus they are ignored when the reduced chi-square is calculated.

The avalanche size difference is defined as $W_n=Z_{i+n}-Z_i$, where $n$ is the temporal interval scale and $Z_i$ is the scale size of the $i{th}$ burst. Here the dimensionless difference is rescaled to
$w_n=W_{n}/\sigma_{W_{n}}$, where $\sigma_{W_{n}}$ is the standard deviation of $W_n$. The Tsallis q-Gaussian function given by \cite{Tsallis1998} is used to fit the PDFs of the size differences, which can be written as
\begin{equation} \label{eq3}
f(w_n)= a[1- b(1-q)w{_n^2}]^{1/1-q},
\end{equation}
where $q$, $a$, and $b$ are free parameters and correspond to the sharpness of the peak, the normalization factor, and the width of the peak, respectively. Note that $q$ describes the departure from the Gaussian function. In other words, when $q\rightarrow 1$, the $q$-Gaussian form reduces to the Gaussian function with mean $\mu=0$ and standard deviation $\sigma=1/\sqrt[]{2b}$. Here, due to the limited
number of data points, we use the cumulative distribution function (CDF) of Tsallis $q$-Gaussian function to fit the size difference distributions \citep{Sang2022,Wang2023}, which can be written as
\begin{equation} \label{eq4}
F(w{_n})=\int_{-\infty}^{w_n} f(w{_n}) dw{_n}.
\end{equation}
The best-fit results are obtained by minimizing the $\chi^2$ statistics.
\section{Results} \label{sec:results}

\subsection{SOC signatures in precursors} \label{sec:results}

\begin{figure*}
\centering
\gridline{
\fig{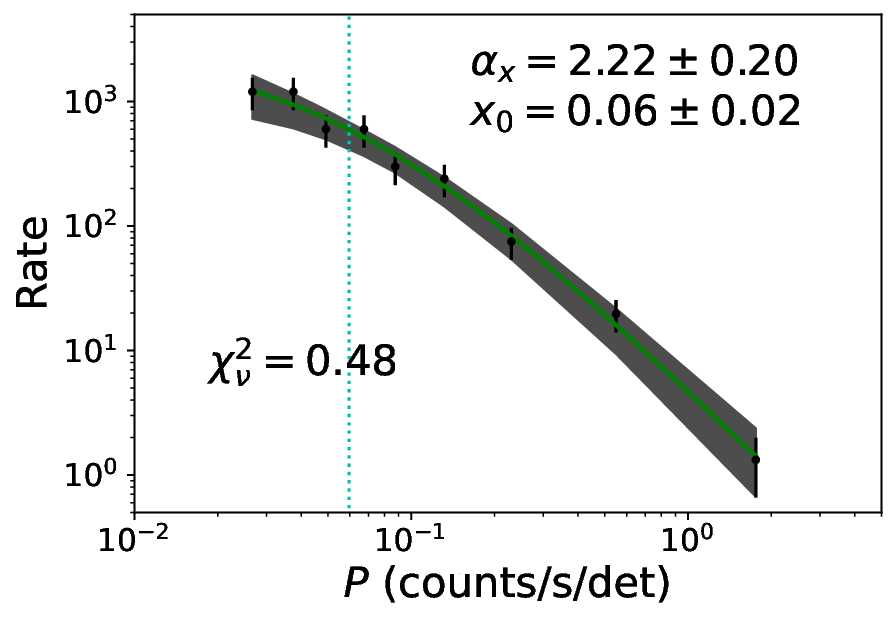}{0.245\textwidth}{(a1)}
 \fig{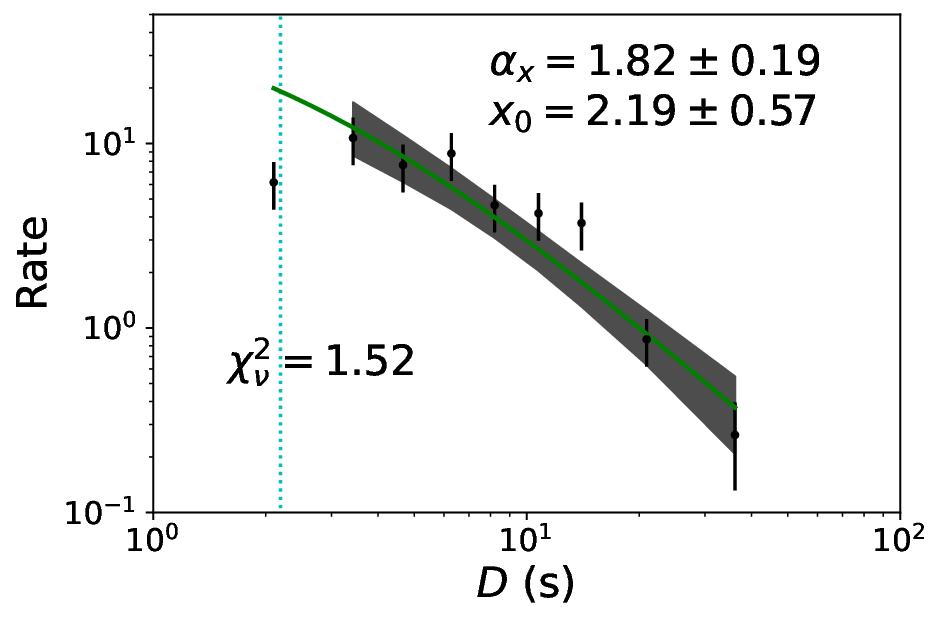}{0.255\linewidth}{(b1)}
 \fig{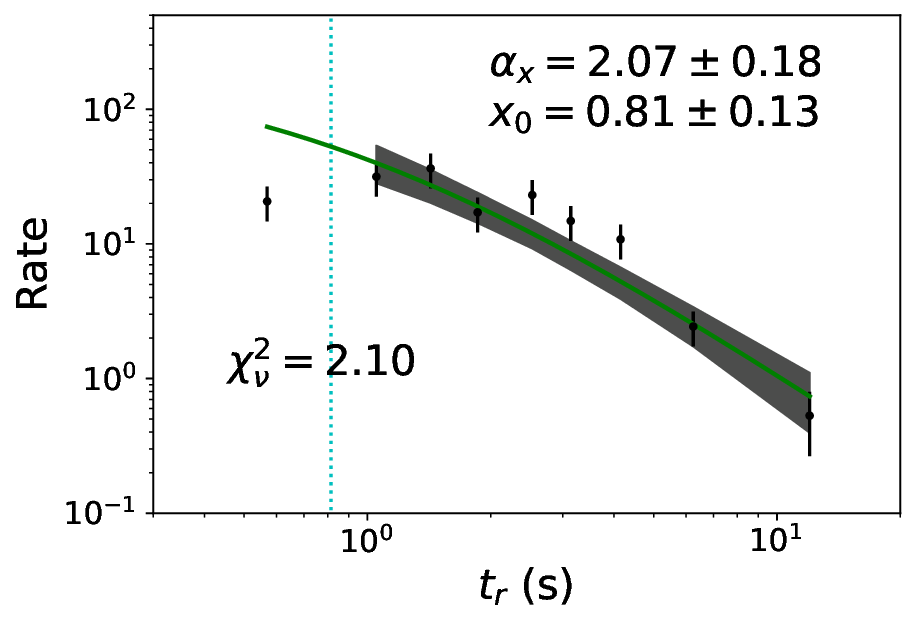}{0.245\textwidth}{(c1)}
 \fig{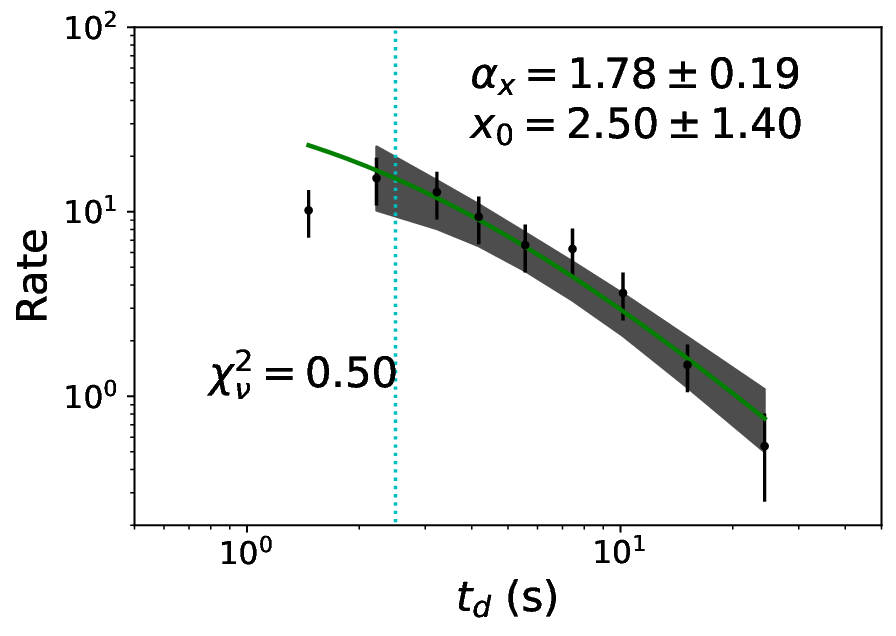}{0.245\textwidth}{(d1)}
          }
 \centering
\gridline{
  \fig{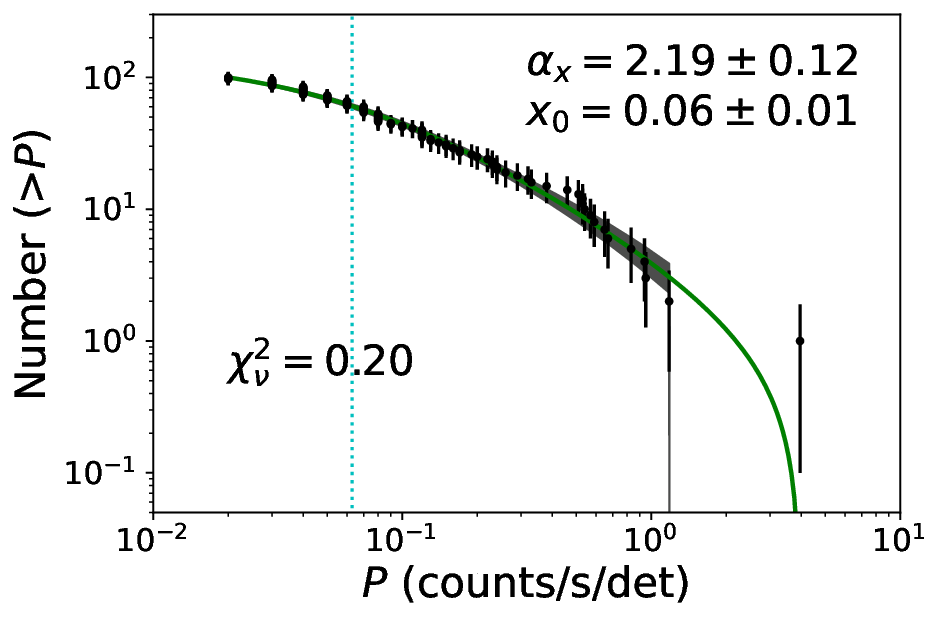}{0.25\textwidth}{(a2)}
\fig{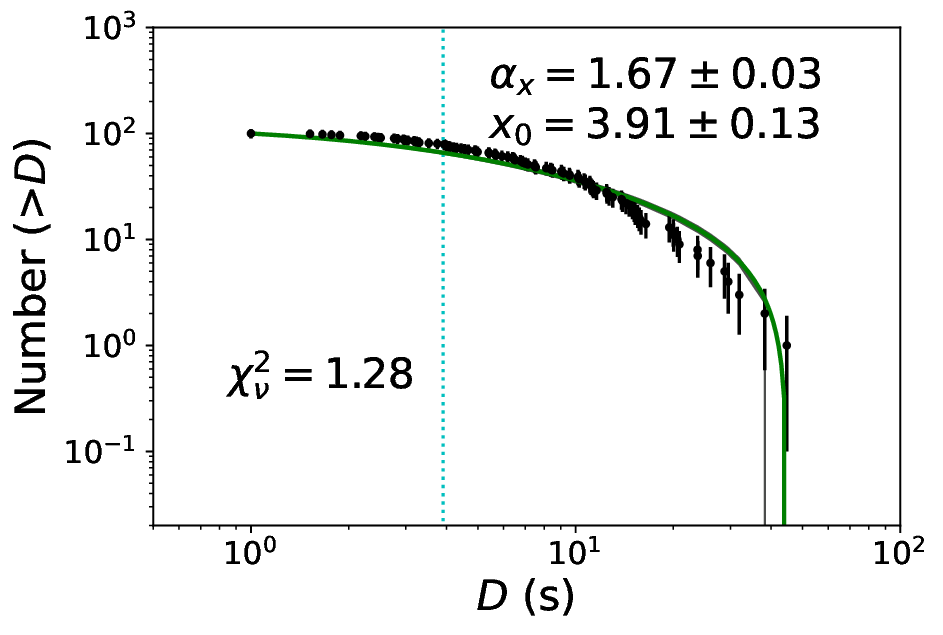}{0.25\textwidth}{(b2)}
 \fig{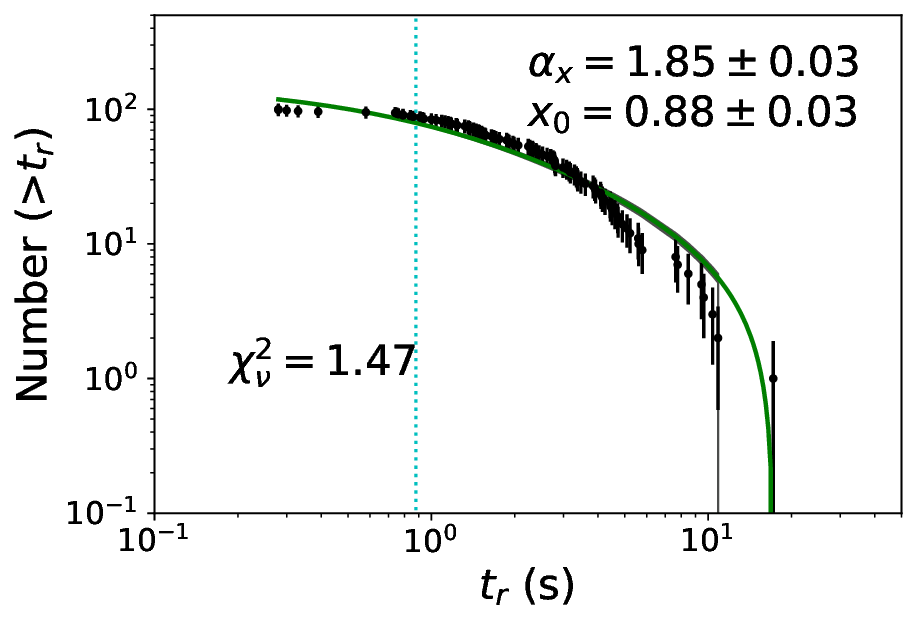}{0.25\textwidth}{(c2)}
 \fig{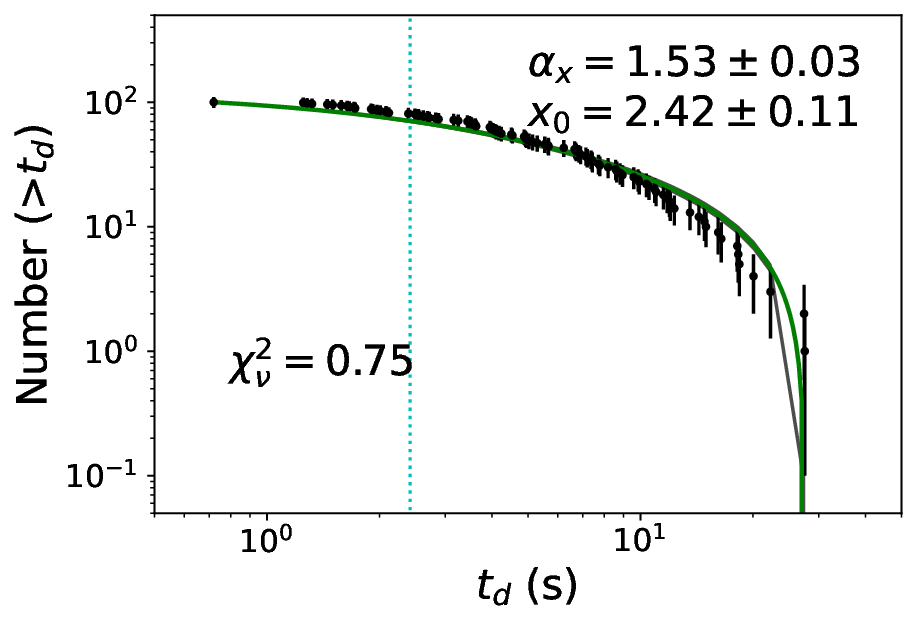}{0.25\textwidth}{(d2)}
          }

\caption{The size distributions of precursors (upper panel: differential distributions, lower panel: cumulative distributions). Panels a - d display the size distributions of peak fluxes ($P$), durations ($D$), rise times ($t_r$), and decay times ($t_d$), respectively. The green solid line is the best fit and the gray region represents the 95\% confidence level, and the dotted line is marked as the threshold $x_{\rm 0}$. \label{pre}}
\end{figure*}

Figure~\ref{pre} shows the fitting results of size distributions for precursors. The differential and cumulative distributions of peak fluxes, durations, rise times, and decay times are shown sequentially in panels (a) - (d). From Panels (a1) - (d1), we find that all the sizes of precursors show the similar power-law distributions. Note that a similar rank-order plot to Paper I is adopted to get the differential distributions. For cumulative distributions, the similar results are shown in Panels (a2) - (d2). The detailed fitting results are listed in Table~\ref{tab1}. The mean values of $\alpha$ of differential and cumulative distributions of peak fluxes, durations, rise times, and decay times are
$\alpha_{mean}$ = 2.21 $\pm$ 0.12, 1.75 $\pm$ 0.10, 1.96 $\pm$ 0.09, and 1.66 $\pm$ 0.10, respectively. As done in Paper I, some noise points below the threshold do
not contribute to the accuracy of the best-fit power-law index and are ignored. These fitting results of power-law differential and cumulative distributions provide strong SOC evidences in precursors.

\begin{figure*}
\centering
\gridline{
  \fig{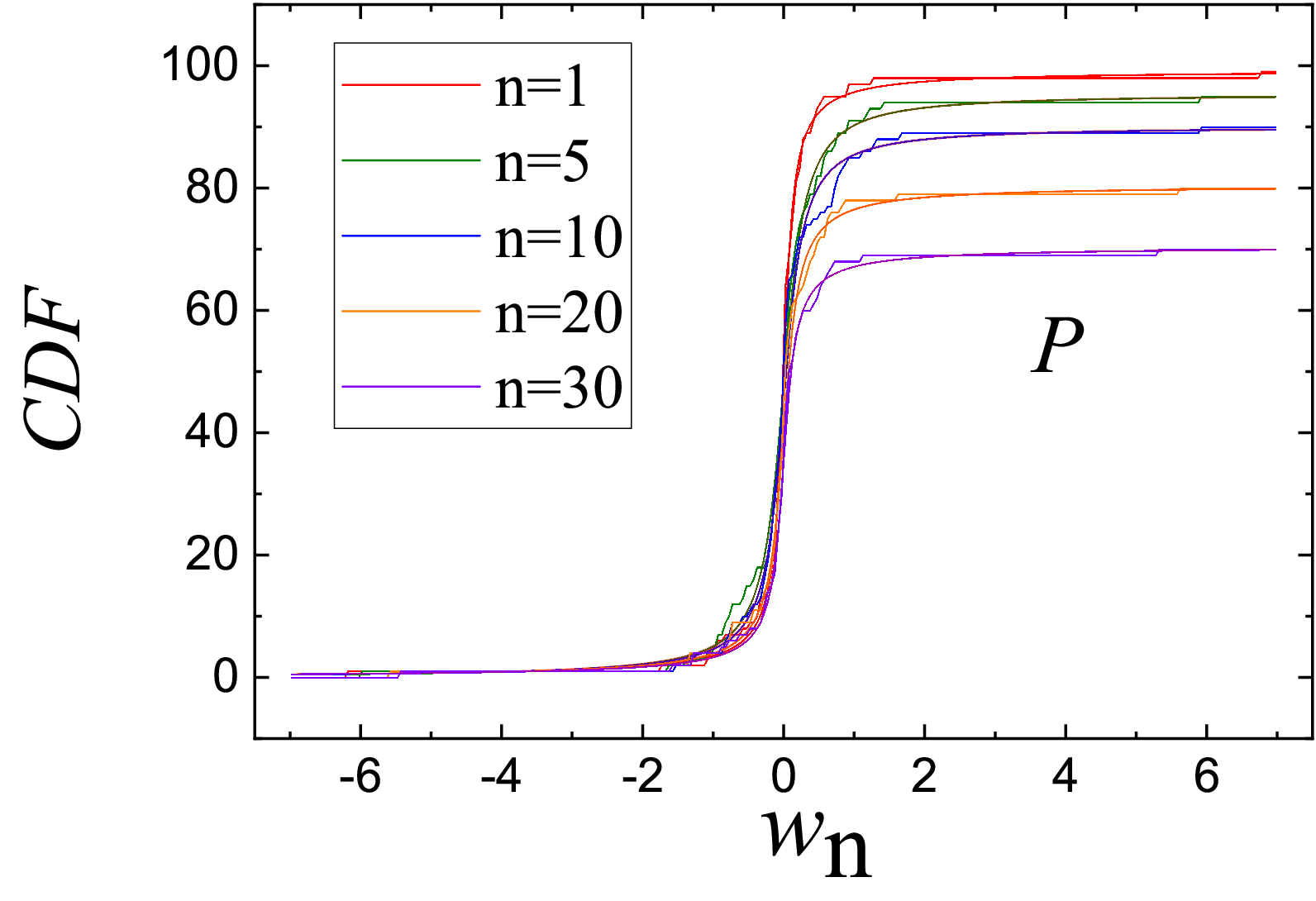}{0.25\textwidth}{(a)}
\fig{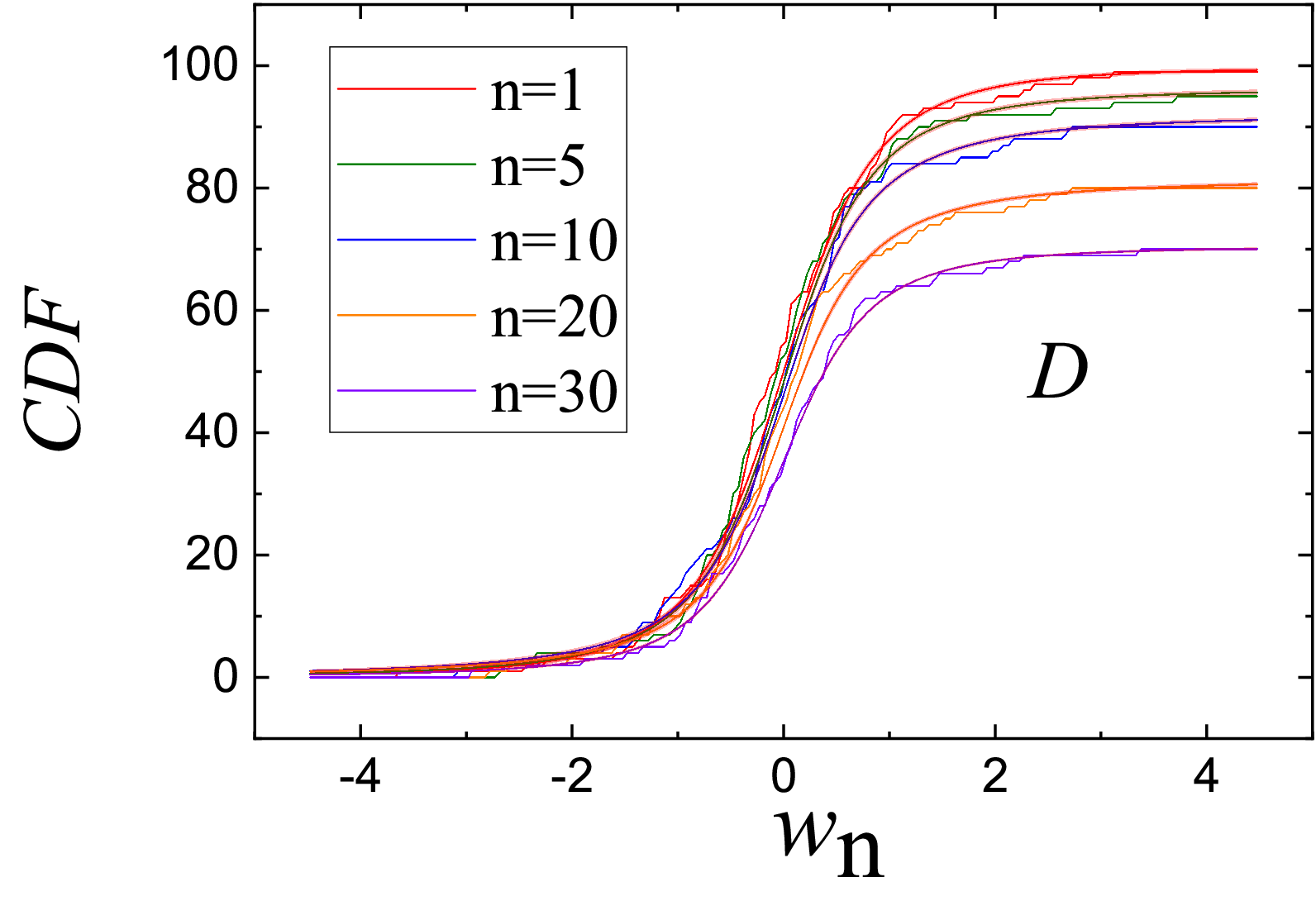}{0.25\textwidth}{(b)}
 \fig{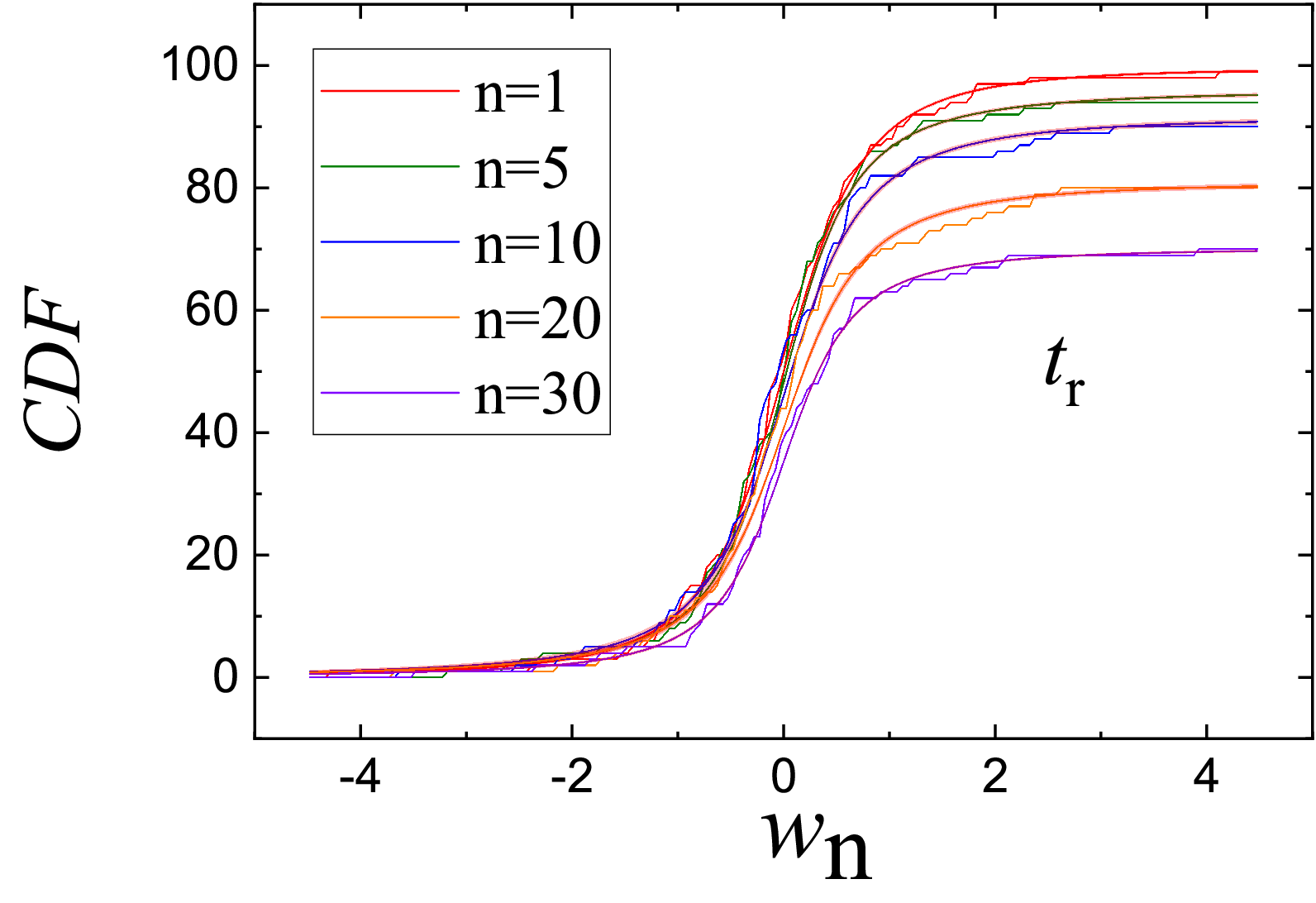}{0.25\textwidth}{(c)}
 \fig{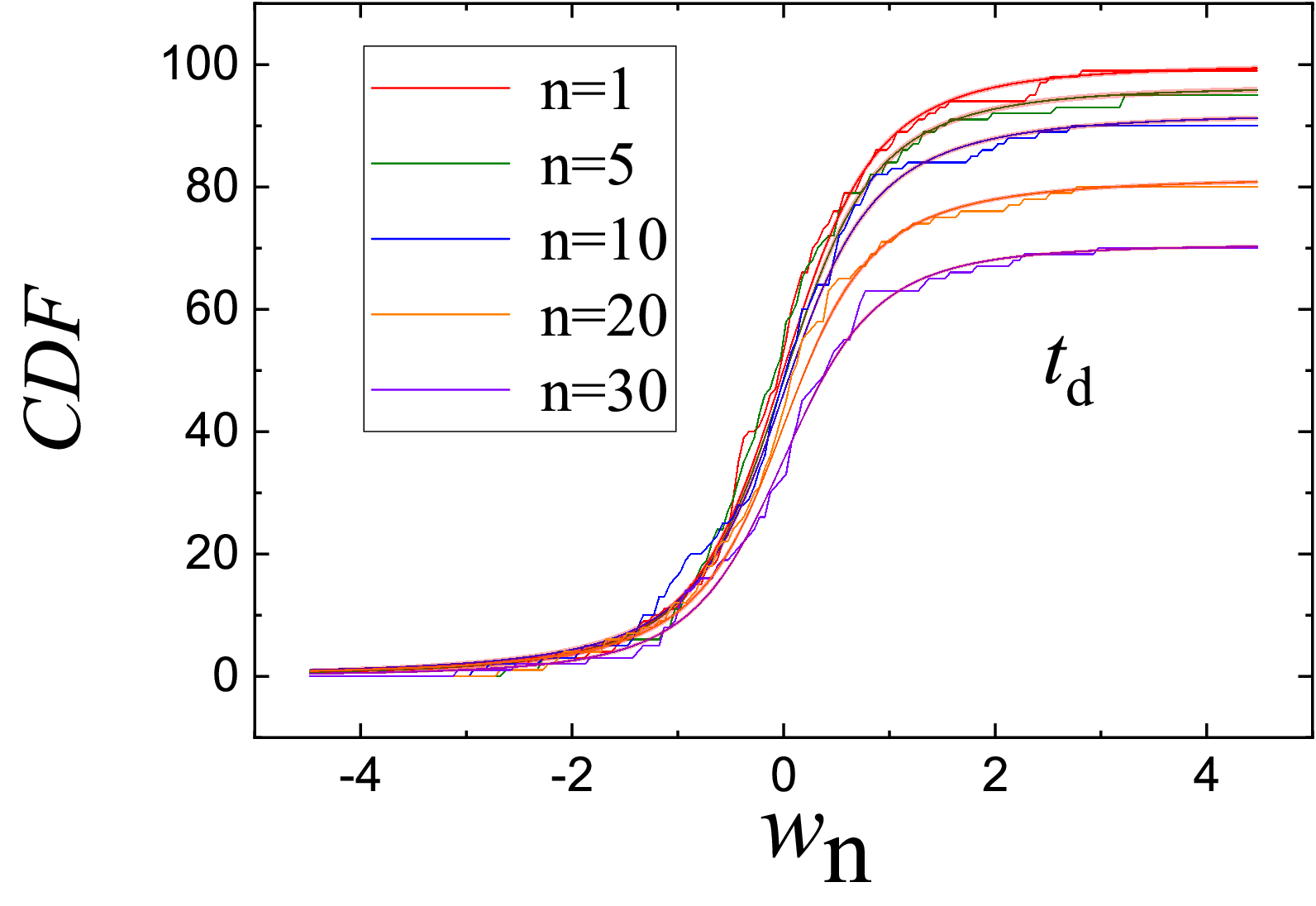}{0.25\textwidth}{(d)}}
 \gridline{
 \fig{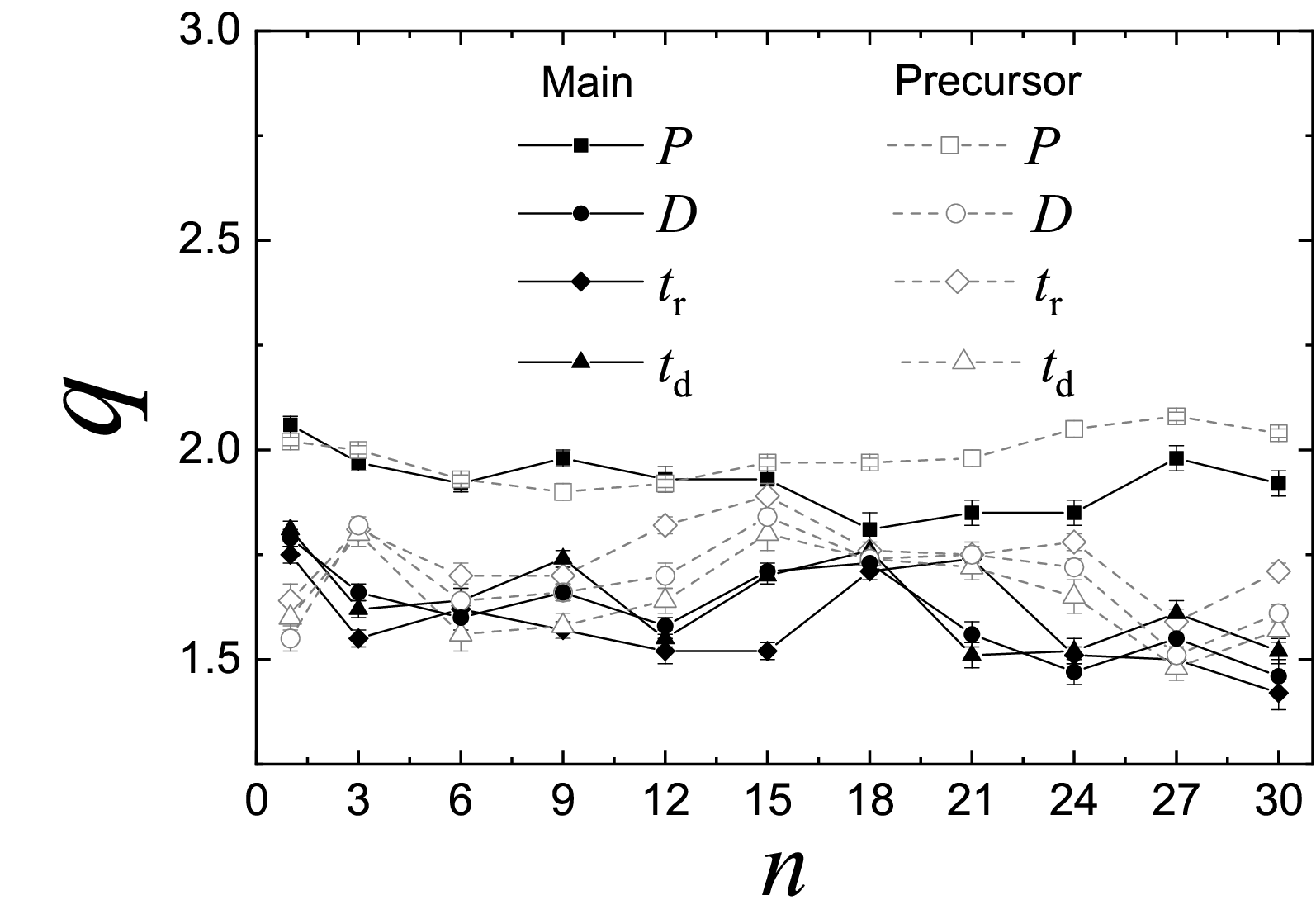}{0.4\textwidth}{(e)}
            }
\caption{The CDFs of the size differences of precursors. Panels (a) - (d) show the size difference distributions of peak fluxes ($P$), durations ($D$), rise times ($t_r$), and decay times ($t_d$) for $n = 1$ (red), $n = 5$ (green), $n = 10$ (blue), $n = 20$ (orange), and $n = 30$ (purple), respectively. The shadow regions represent the 95\% confidence level, the smooth lines are the best fits. Panel (e) shows the best-fit values of $q$ in the $q$-Gaussian distributions as a function of $n$ for the precursors (the open symbols) and the main bursts (the filled symbols). \label{pre2}}
\end{figure*}

Figure~\ref{pre2} shows the CDFs of avalanche size difference of peak fluxes, durations, rise times, and decay times, peak-time intervals, and quiescent times with interval scale $n = 1, 5, 10, 20, 30$. We can see that the observation data can be fitted very well by the $q$-Gaussian model.
Furthermore, we investigate the scale invariance properties in the interval $n$. We fit the CDFs of avalanche size differences of peak fluxes, durations, rise times, and decay times for $n$ from 1, 3 to 30 every three with Function~\ref{eq4}. The mean values of $q$ are $q_{mean}$ = 1.99 $\pm$ 0.01, 1.69 $\pm$ 0.01, 1.74 $\pm$ 0.01, and 1.65 $\pm$ 0.01, respectively. The $q$ values as a function of $n$ are plotted in the panel (e) of Figure 2. We can see that the $q$ values are independence on $n$ and approximately constant for different $n$, confirming that the scale-invariant properties do indeed exist in precursors.

\subsection{Comparison with main bursts} \label{sec:results}
\begin{figure*}
\centering
\gridline{
 \fig{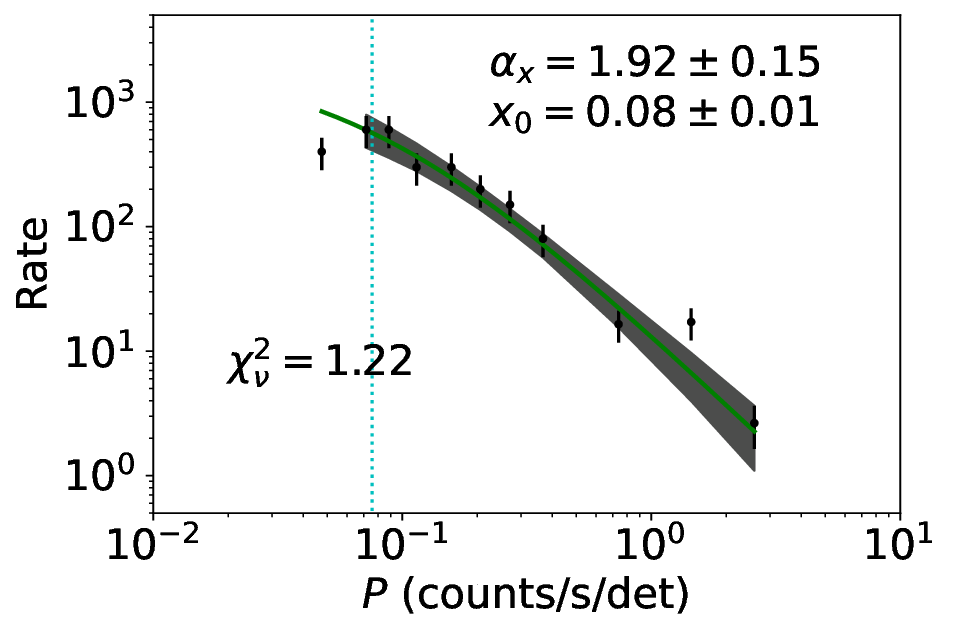}{0.26\linewidth}{(a1)}
 \fig{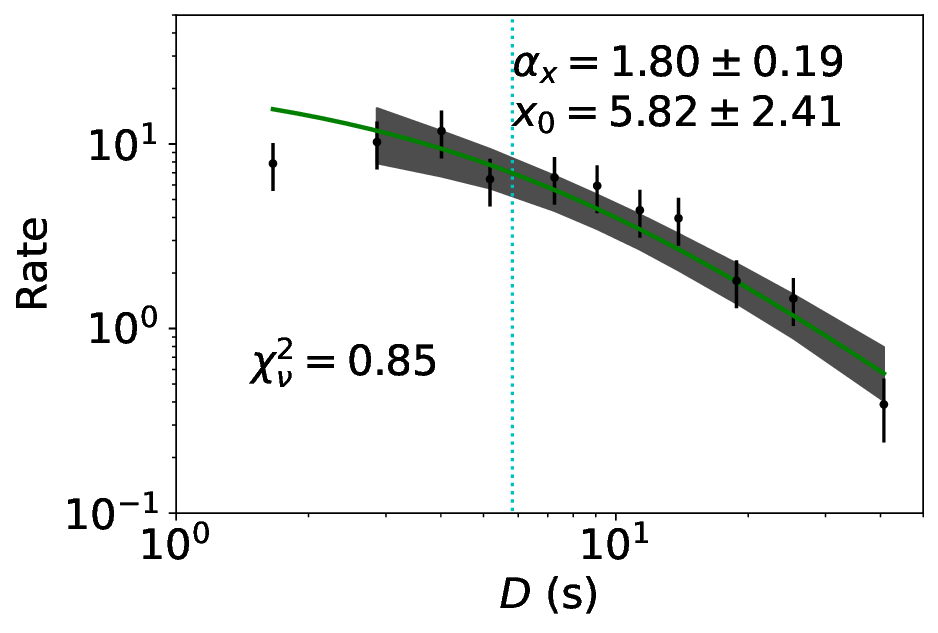}{0.25\textwidth}{(b1)}
 \fig{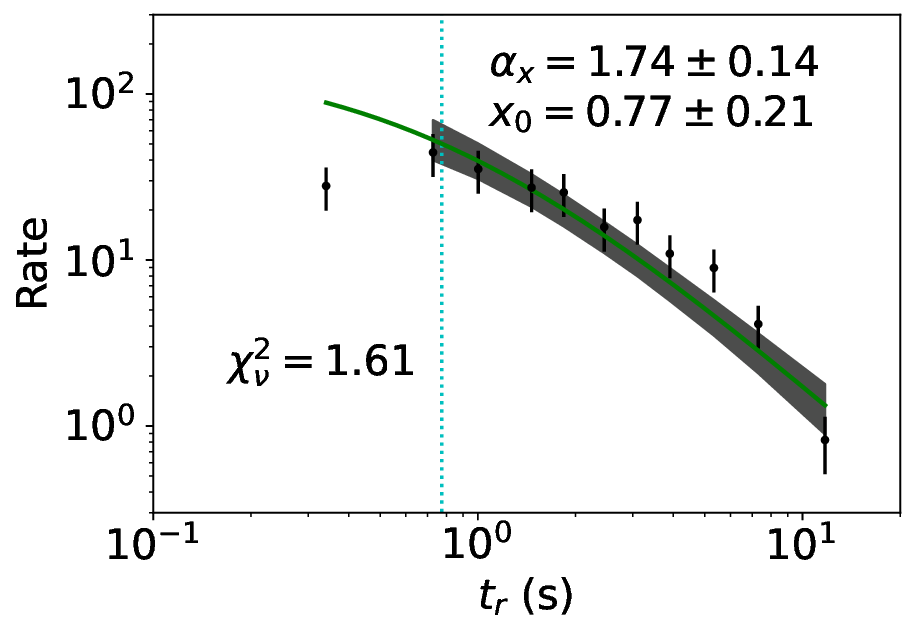}{0.245\textwidth}{(c1)}
 \fig{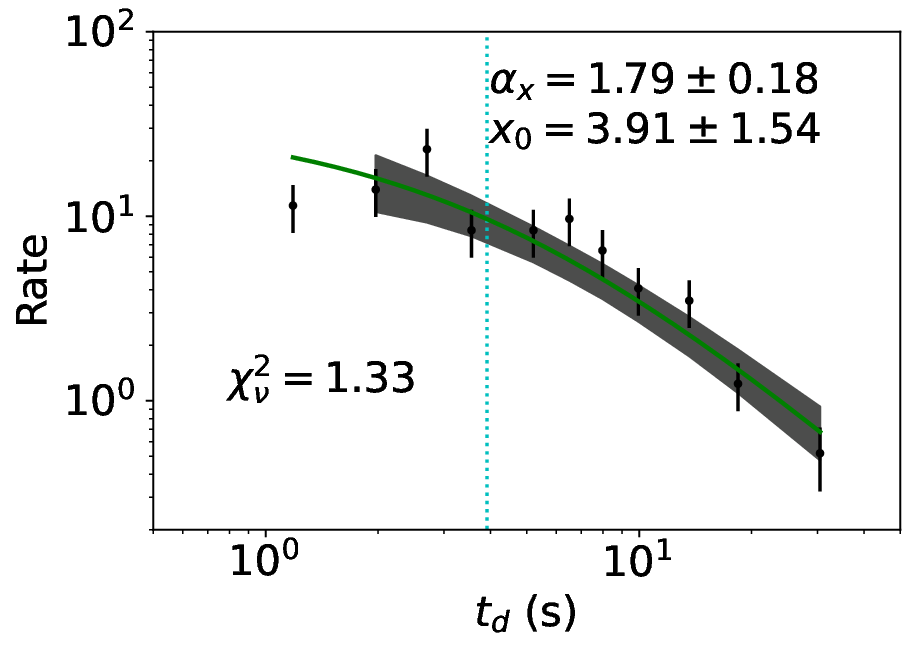}{0.245\textwidth}{(d1)}
          }
\centering
\gridline{
\fig{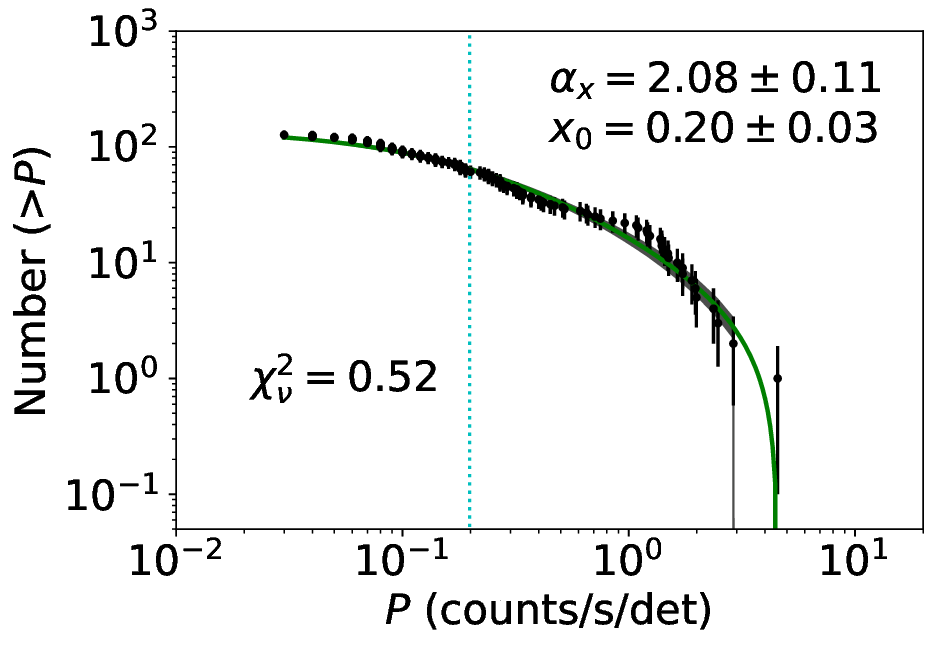}{0.245\textwidth}{(a2)}
  \fig{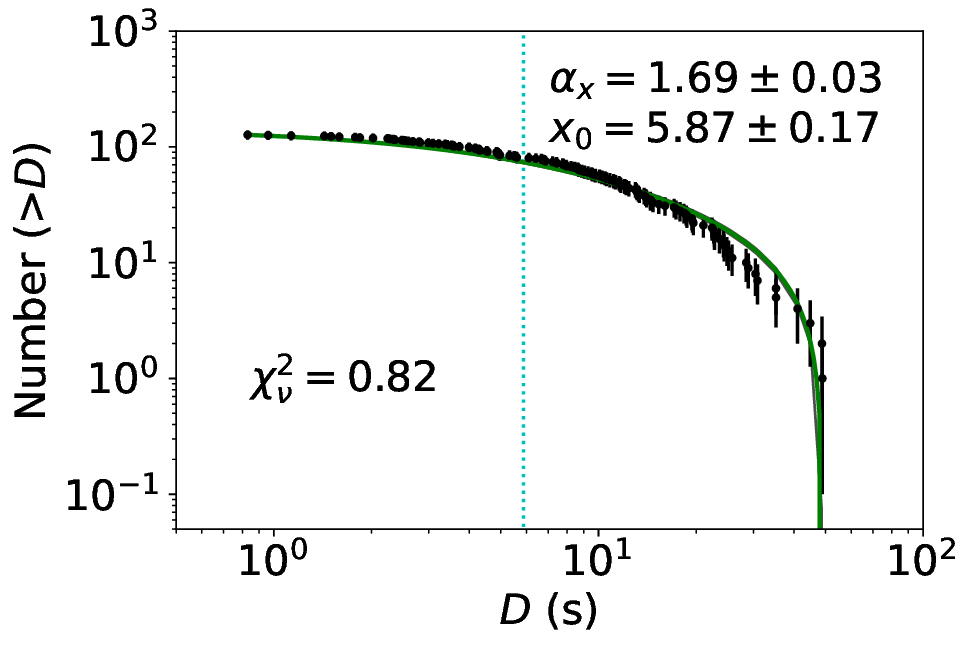}{0.255\textwidth}{(b2)}
 \fig{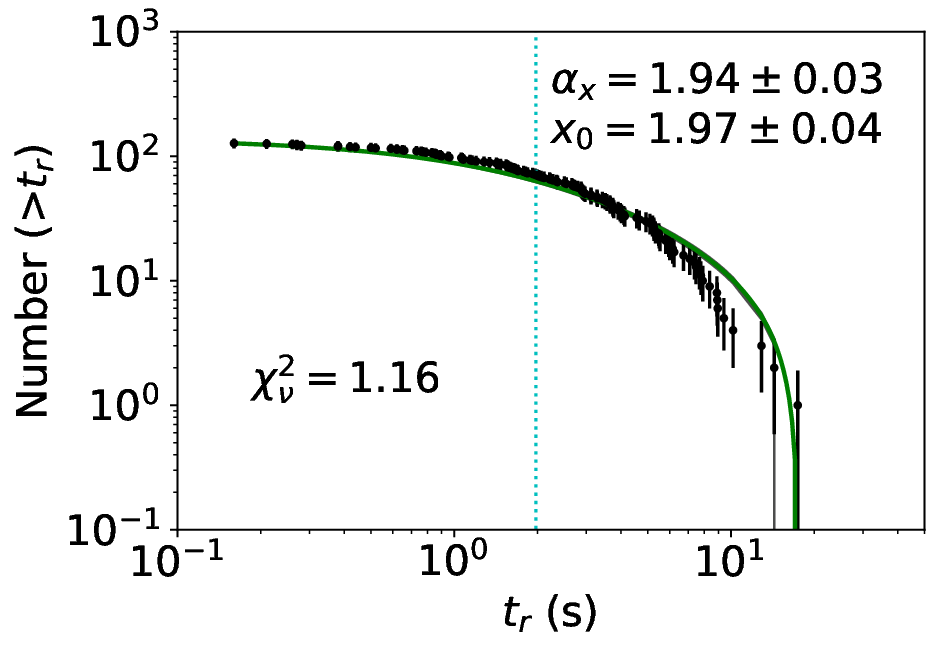}{0.245\textwidth}{(c2)}
 \fig{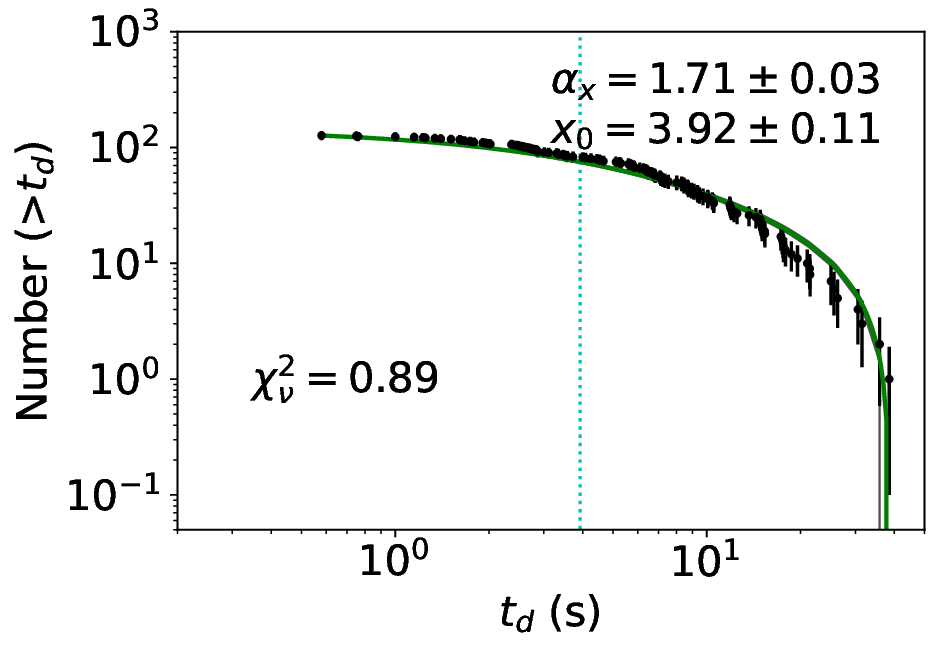}{0.245\textwidth}{(d2)}
          }

\caption{The size distributions of main bursts. The symbols are the same as those in Figure 1. \label{main}}
\end{figure*}

In order to reveal the potential connection between precursors and main bursts, the similar studies are made in main bursts.  The size distributions of peak flux, durations, rise times, and decay times of main bursts are illustrated sequentially in panels (a) - (d) of Figure~\ref{main}. The detailed fitting results are presented in Table~\ref{tab1}. It can be seen that the results of main bursts tend to be similar to those of the precursors. In other words, the signatures of the SOC behaviors in main bursts are also legible, which are essentially consistent with the GRBs without precursors \citep{Lv2020,Lxj6}. The mean values of $\alpha$ of differential and cumulative distributions of peak fluxes, durations, rise times, and decay times are $\alpha_{mean}$ = 2.00 $\pm$ 0.09, 1.75 $\pm$ 0.10, 1.84 $\pm$ 0.07, and 1.75 $\pm$ 0.09, respectively. These values are in good agreement with the results given by Paper I for the BATSE and Swift short GRBs \citep{Lxj6}.

\begin{figure*}
\centering
\gridline{
\fig{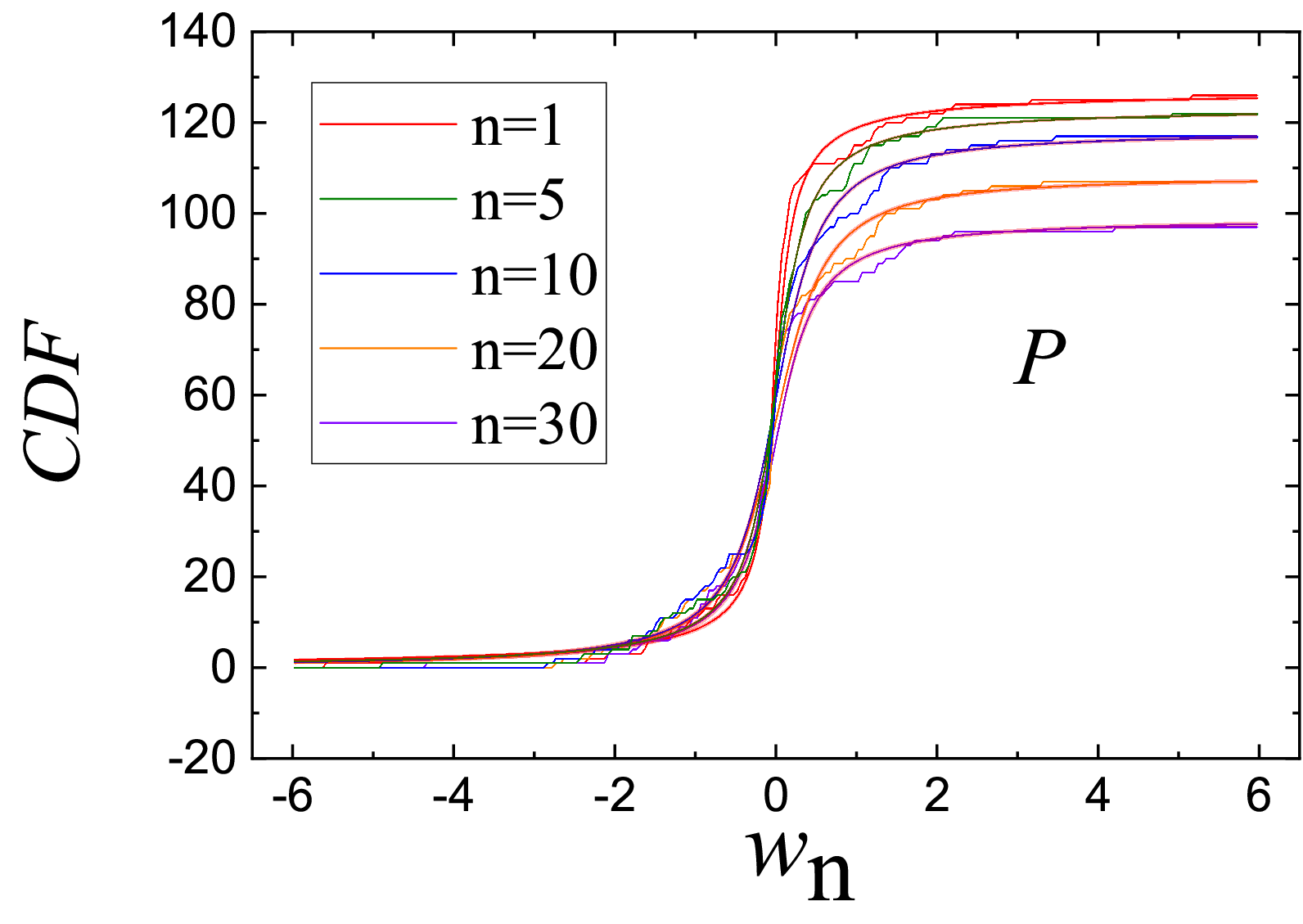}{0.25\textwidth}{(a)}
\fig{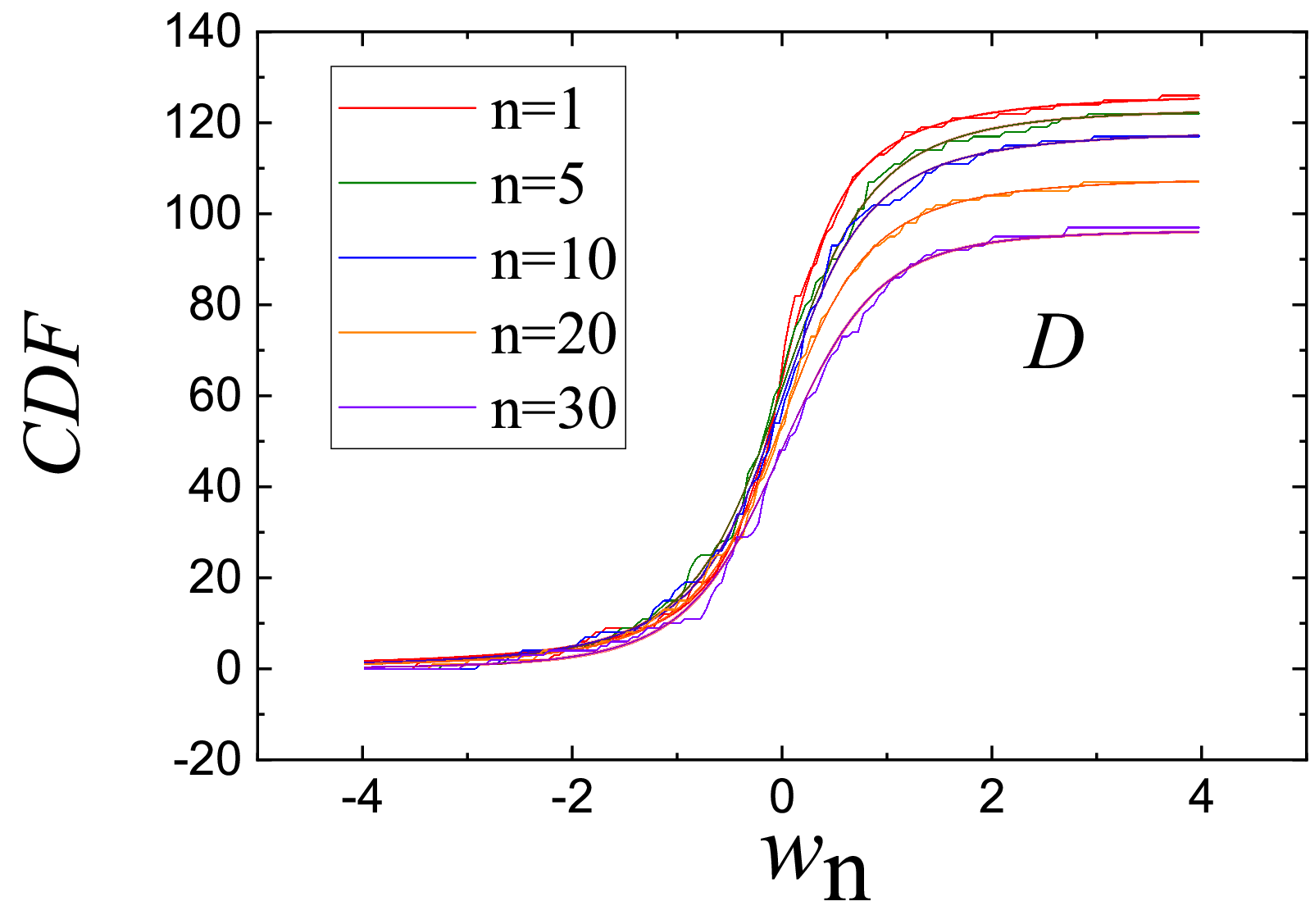}{0.25\textwidth}{(b)}
\fig{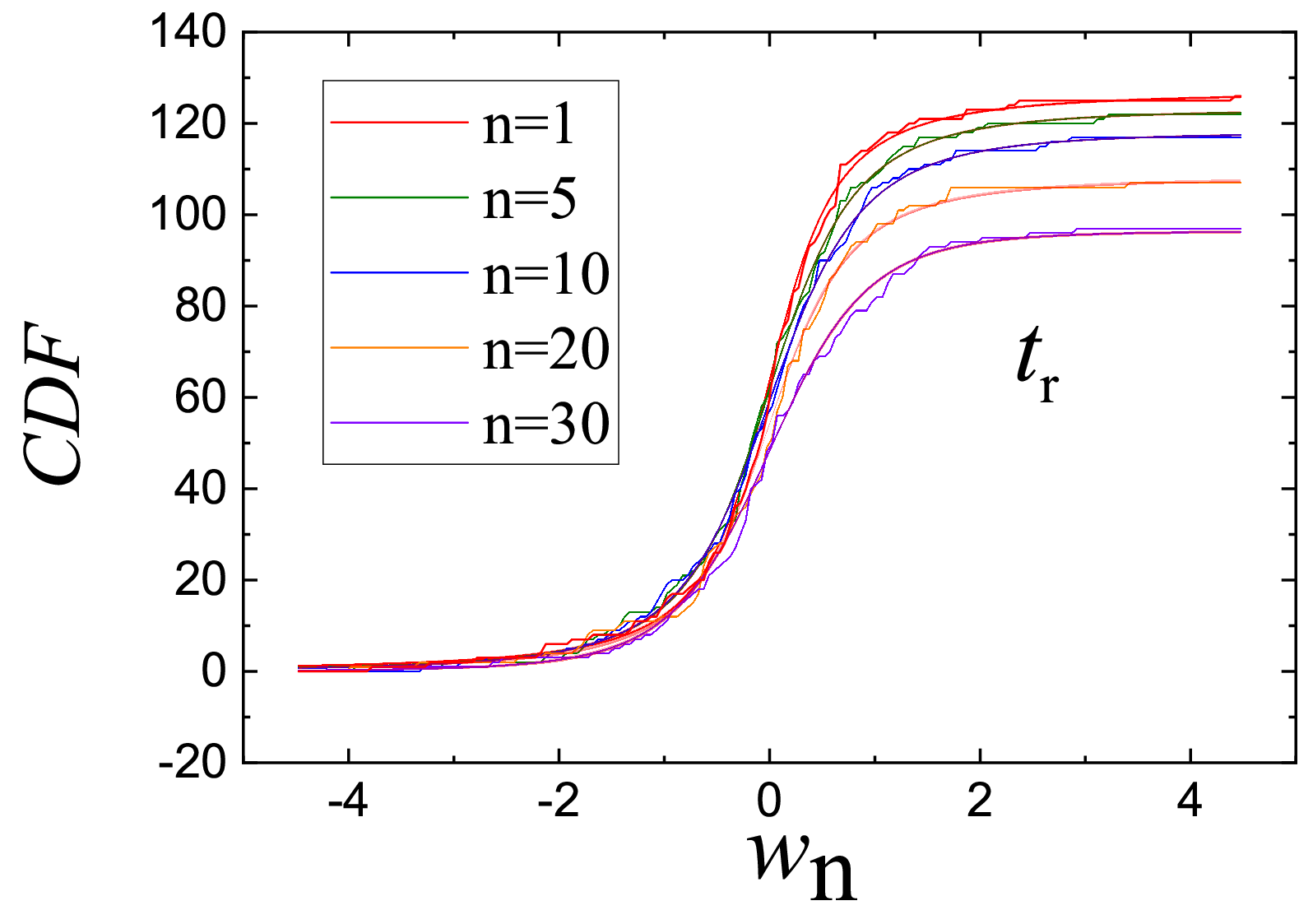}{0.25\textwidth}{(c)}
\fig{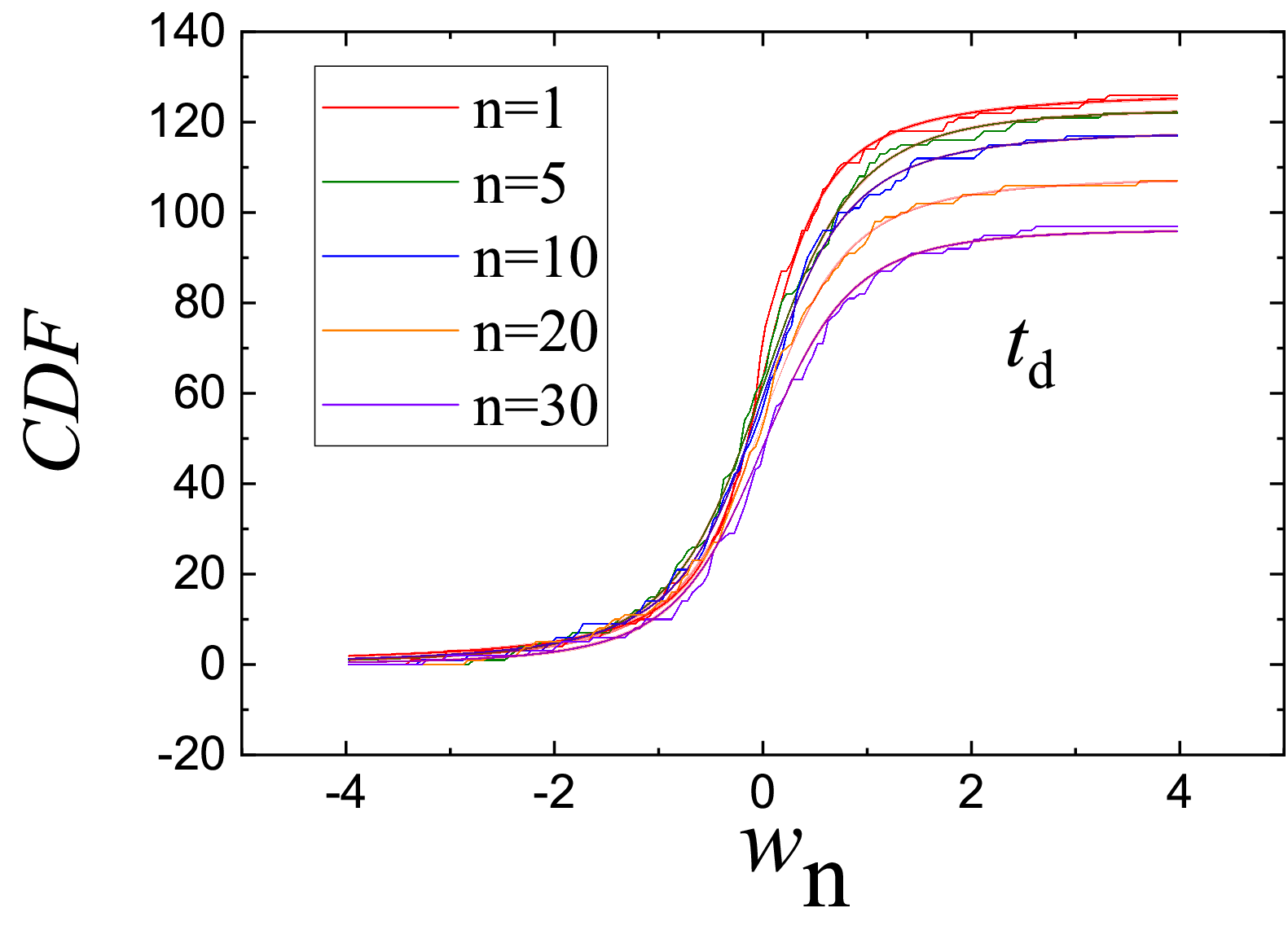}{0.25\textwidth}{(d)}}
\caption{The CDFs of the size differences of main bursts. The symbols are the same as those in Figure 2. \label{main2}}
\end{figure*}

Similar scale-invariant behaviors of main bursts are shown in Figure~\ref{main2}. Panels (a) - (d) show the CDFs of avalanche size difference of peak fluxes, durations, rise times, and decay times with interval scale $n = 1, 5, 10, 20, 30$. For comparison, we fit the CDFs of size differences of precursors for $n$ from 1, 3 to 30 every three and the results are also displayed in Panel (e) of Figure 2. The mean values of $q$ of peak fluxes, durations, rise times, and decay times for main bursts are $q_{mean}$ = 1.93 $\pm$ 0.01, 1.62 $\pm$ 0.01, 1.58 $\pm$ 0.01, and 1.63 $\pm$ 0.01, respectively. Obviously, the independence of the $q$ values on $n$ for main bursts still remains, which is similar to that of the precursors.

\begin{figure*}
\centering
 \gridline{
 \fig{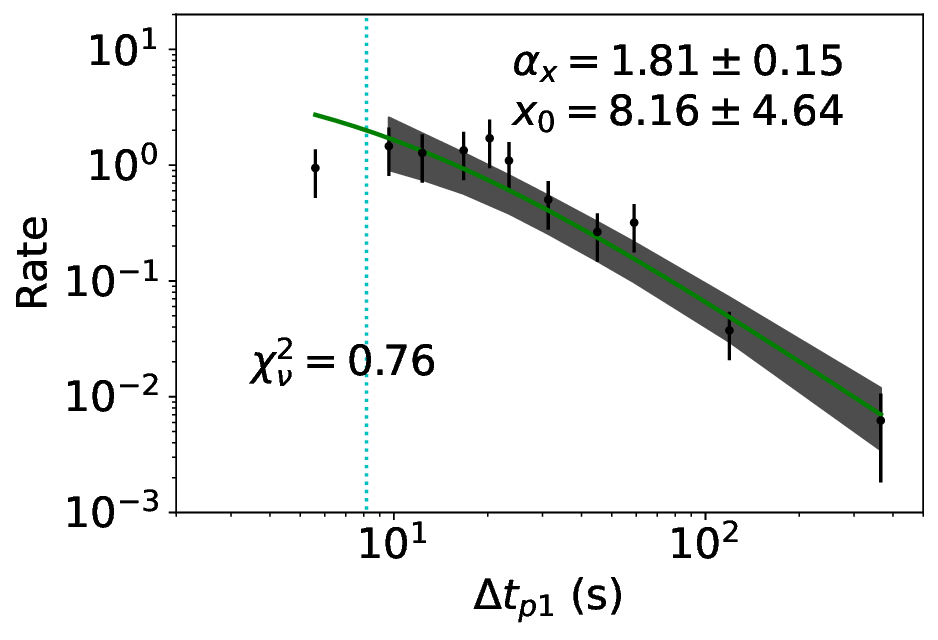}{0.245\textwidth}{(a1)}
  \fig{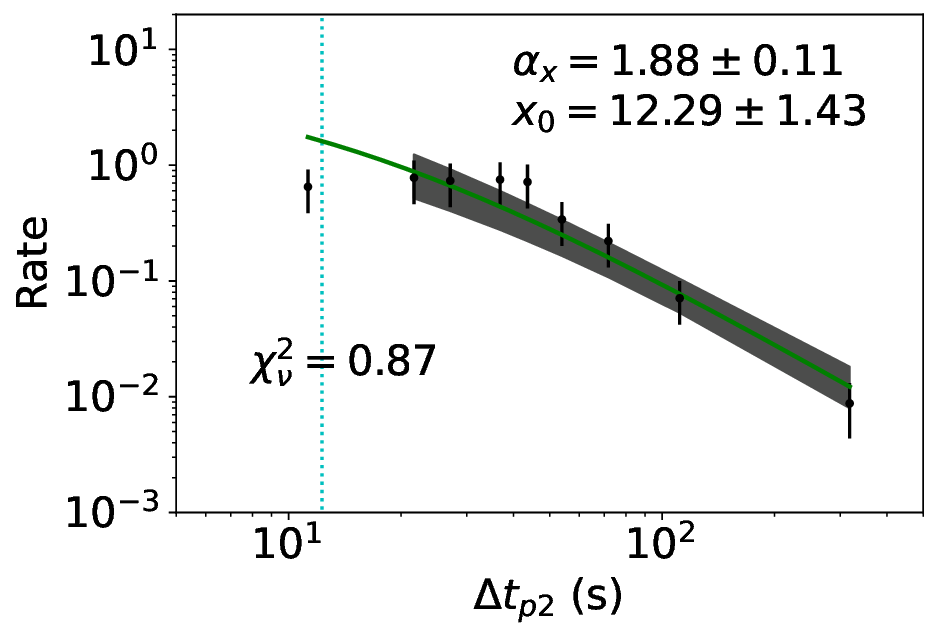}{0.245\textwidth}{(b1)}
   \fig{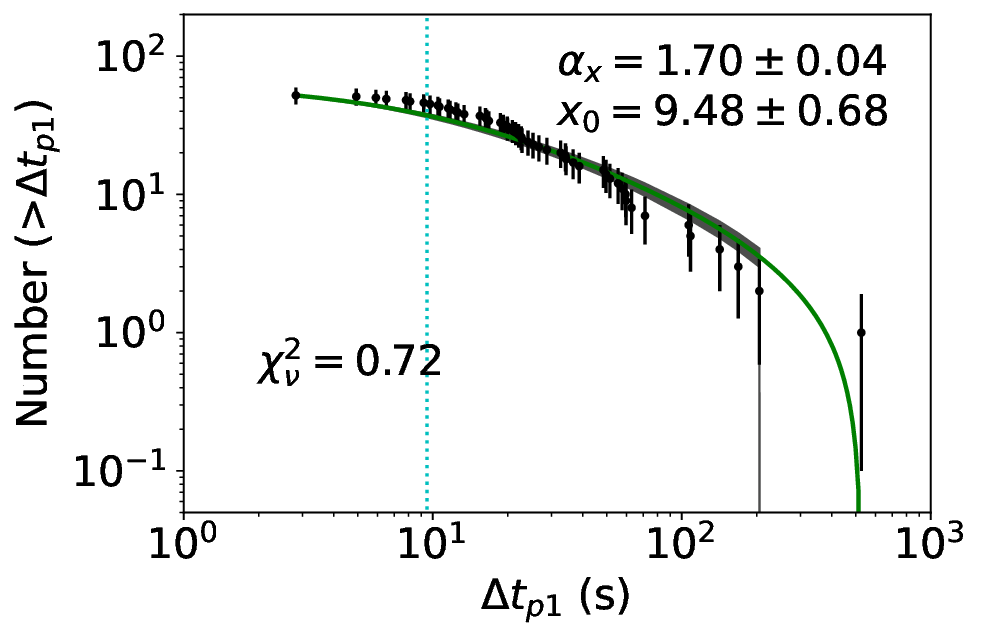}{0.255\textwidth}{(c1)}
  \fig{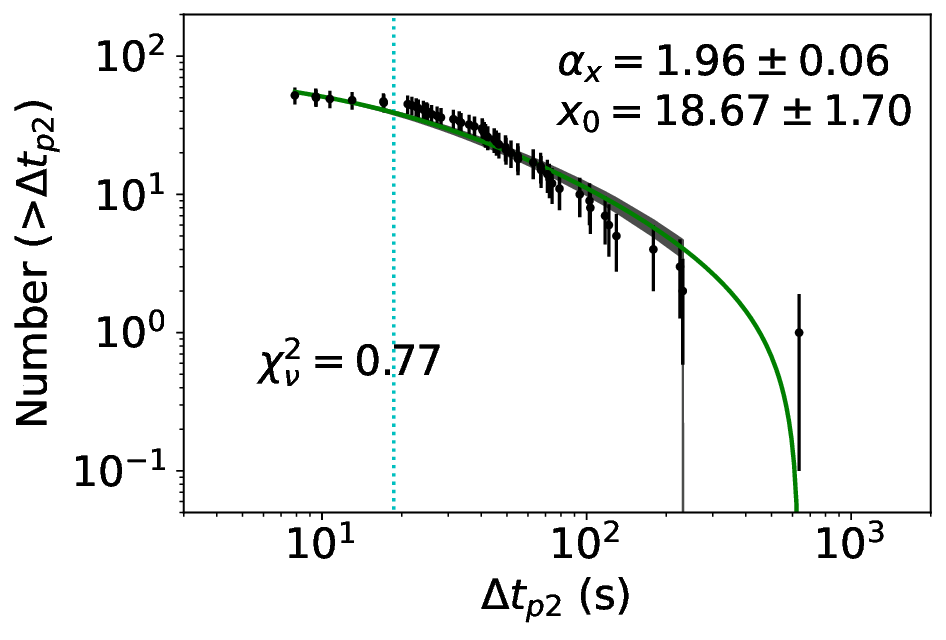}{0.245\textwidth}{(d1)}
 }
\gridline{
\fig{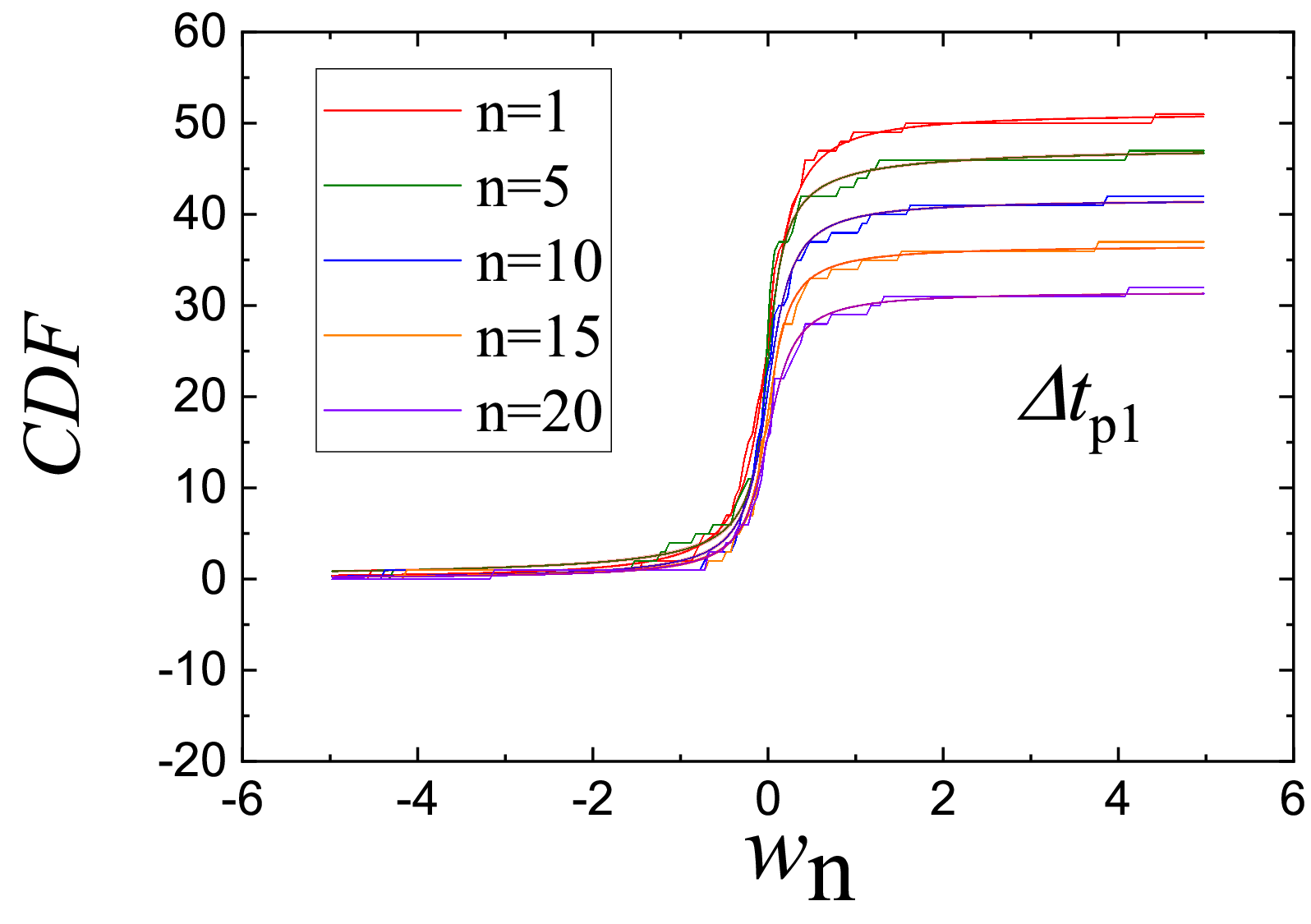}{0.3\textwidth}{(a2)}
\fig{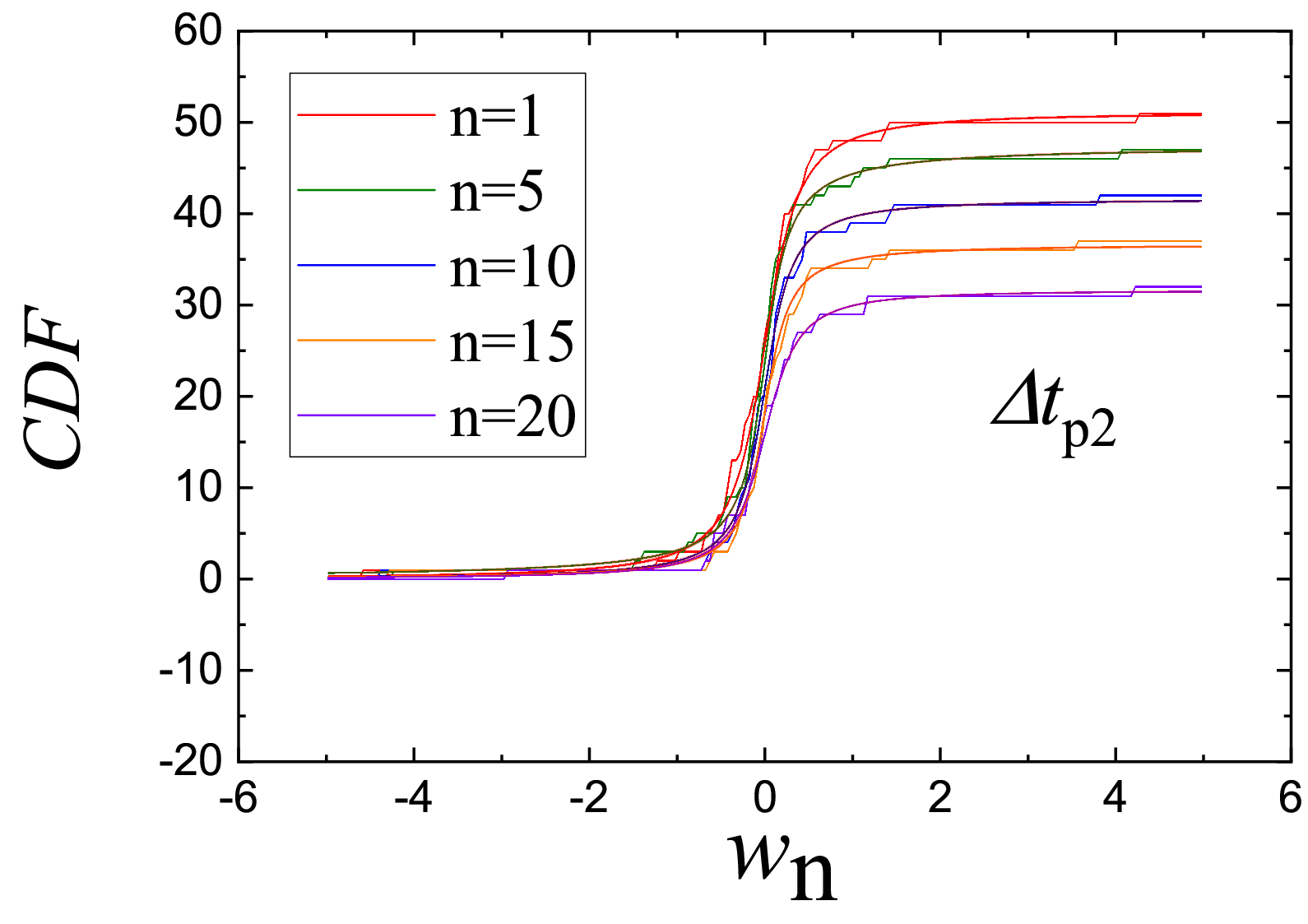}{0.3\textwidth}{(b2)}
\fig{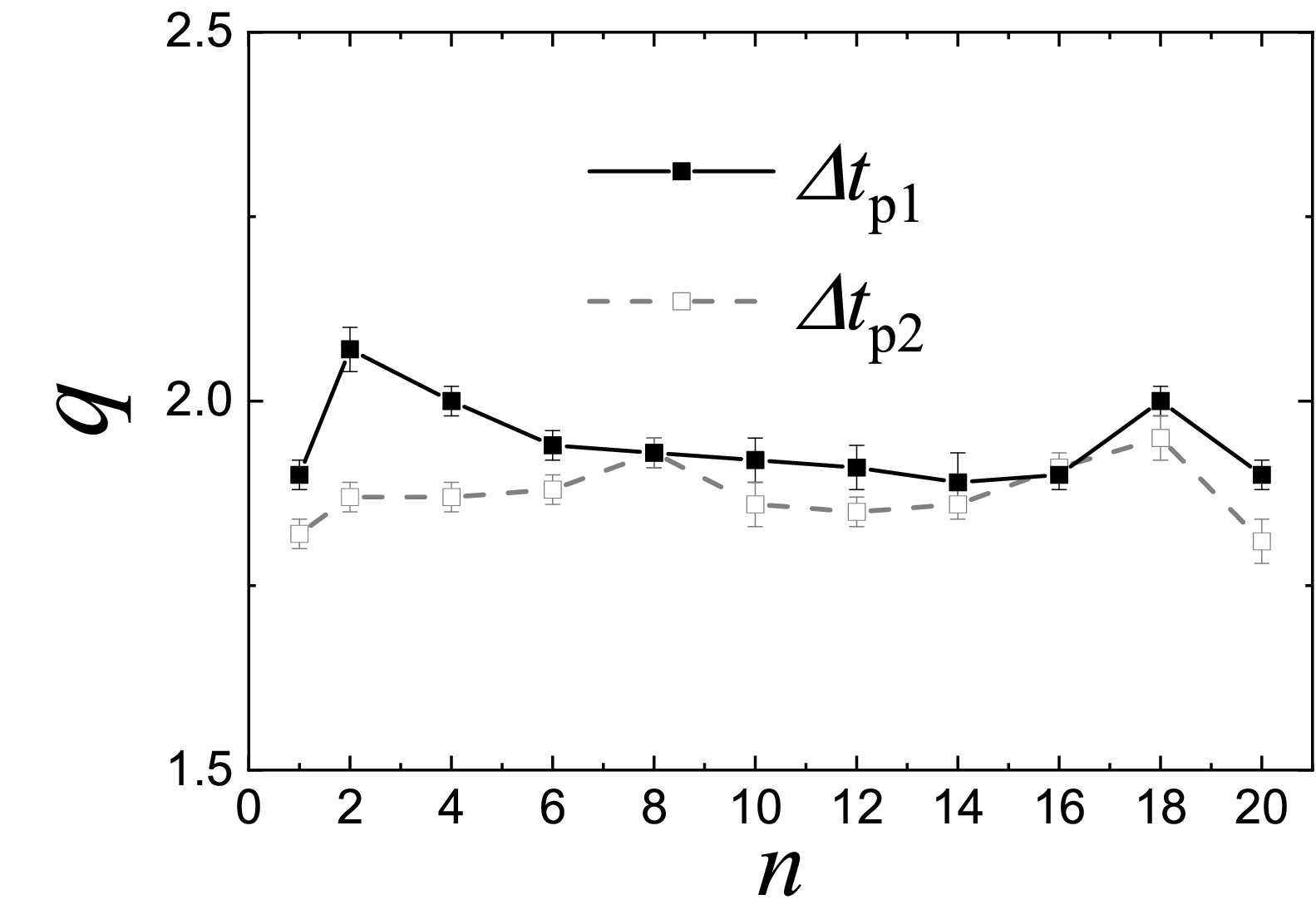}{0.4\textwidth}{(e)}
            }
\caption{The distributions of two kinds of waiting times between precursors and main bursts (upper panel: differential and cumulative size distributions, lower panel: size difference distributions). Panels (a2) and (b2) show the size difference distributions of peak-time intervals ($\Delta t_ {\rm p1}$) and quiescent times ($\Delta t_ {\rm p2}$) for $n = 1$ (red), $n = 5$ (green), $n = 10$ (blue), $n =15$ (orange), and $n = 20$ (purple), respectively. The shadow regions represent the 95\% confidence level, the smooth lines are the best fits. Panel (e) shows the best-fit values of $q$ in the $q$-Gaussian distributions as a function of $n$ for peak-time intervals (the filled symbols) and quiescent times (the open symbols). \label{tp}}
\end{figure*}

Additionally, we also inspect the size distributions and the scale-invariant behaviors of two kinds of waiting times, including the peak-time intervals and the quiescent times between precursors and main bursts. The similar results are shown in Figure~\ref{tp}. Panels (a2) - (b2) show the CDFs of two kinds of waiting time differences with interval scale $n = 1, 5, 10, 15, 20$. The mean $\alpha$ values of differential and cumulative distributions are $\alpha_{mean}$ = 1.76 $\pm$ 0.08 and 1.92 $\pm$ 0.06, respectively. Considering the smaller sample numbers of these waiting time, we just investigate the CDFs of avalanche size difference from $n$ = 1, 2 to 20 every second. The mean $q$ values are $q_{mean}$ = 1.94 $\pm$ 0.01, and 1.87 $\pm$ 0.01, respectively. It is found that there is no substantial difference with the other scale sizes, indicating the common SOC signatures in the GRB prompt phases.

\begin{table*}[htbp]
\centering
\caption{The best-fitting results of the size distributions and the size difference distributions.  \label{tab1}}
\begin{tabular}{cc|cccc|cc}
\hline
\hline
Components & parameter        &$P$          &$D$      & $t_{\rm r}$     & $t_{\rm d}$    &$\Delta t_ {\rm p1}$&$\Delta t_ {\rm p2}$\\
\hline
Precursor &$\alpha_{\rm d}$  & 2.22 $\pm$ 0.20& 1.82 $\pm$ 0.19& 2.07 $\pm$ 0.18& 1.78 $\pm$ 0.19& 1.81 $\pm$ 0.15& 1.88 $\pm$ 0.11\\
         &$x_{\rm 0d}$  & 0.06 $\pm$ 0.02& 2.19 $\pm$ 0.57& 0.81 $\pm$ 0.13& 2.50 $\pm$ 1.40& 8.16 $\pm$ 4.64 & 12.29 $\pm$ 1.43\\
         &$\alpha_{\rm c}$ &2.19 $\pm$ 0.12 & 1.67 $\pm$ 0.03 &  1.85 $\pm$ 0.03& 1.53 $\pm$ 0.03& 1.70 $\pm$ 0.04  & 1.96 $\pm$ 0.06\\
         &$x_{\rm 0c}$ &0.06 $\pm$ 0.01 & 3.91 $\pm$ 0.13 & 0.88 $\pm$ 0.03& 2.42 $\pm$ 0.11 & 9.48 $\pm$ 0.68 & 18.67 $\pm$ 1.70\\
         &$\alpha_{mean}$ & 2.21 $\pm$ 0.12 & 1.75 $\pm$ 0.10&1.96 $\pm$ 0.09& 1.66 $\pm$ 0.10&1.76 $\pm$ 0.08 & 1.92 $\pm$ 0.06\\
         &$q_{mean}$ &1.99 $\pm$ 0.01 &1.69 $\pm$ 0.01&1.74 $\pm$ 0.01 &1.65 $\pm$ 0.01 &1.94 $\pm$ 0.01 &1.87 $\pm$ 0.01 \\
\hline
Main   &$\alpha_{\rm d}$& 1.92 $\pm$ 0.15 & 1.80 $\pm$ 0.19& 1.74 $\pm$ 0.14& 1.79 $\pm$ 0.18 & \\
       &$x_{\rm 0d}$ & 0.08 $\pm$ 0.01& 5.82 $\pm$ 2.41& 0.77 $\pm$ 0.21& 3.91 $\pm$ 1.54& \\
       &$\alpha_{\rm c}$ & 2.08 $\pm$ 0.11& 1.69 $\pm$ 0.03& 1.94 $\pm$ 0.03& 1.71 $\pm$ 0.03& &\\
       &$x_{\rm 0c}$ & 0.20 $\pm$ 0.03&5.87 $\pm$ 0.17& 1.97 $\pm$ 0.04& 3.92 $\pm$ 0.11& &\\
       &$\alpha_{mean}$ &2.00 $\pm$ 0.09& 1.75 $\pm$ 0.10& 1.84 $\pm$ 0.07& 1.75 $\pm$ 0.09 &&\\
       &$q_{mean}$ &1.93 $\pm$ 0.01 &1.62 $\pm$ 0.01&1.58 $\pm$ 0.01 &1.63 $\pm$ 0.01 \\
\hline\hline
\end{tabular}
\end{table*}



\section{Discussions} \label{sec:result3}
According to the prediction proposed by \cite{Aschwanden2012,Aschwanden2014} for a fractal-diffusive SOC model, the indexes $\alpha_P$ and $\alpha_T$ of the power-law size distributions for the peak flux and duration can be derived from $\alpha_P = 2-1/S $ and $\alpha_T = (1+S)\beta/2$, where $S = 1, 2, 3$ are the spatial dimensions of the SOC model and $\beta = 1 $ is the diffusive spreading exponent for the classical diffusion. Obviously,
the spatial dimension $S = 3$ can be obtained to explain the SOC behavior of precursors by taking the mean values of the power-law indexes. Likewise, the same spatial dimension $S = 3$ can be determined by the quite same mean values of the power-law indexes of the main bursts, indicating that precursors and main bursts might share the similar physical processes. Moreover, we find that the results of main bursts are consistent with those reported by our Paper I \citep{Lxj6} and \cite{Lv2020} in despite of different GRB samples, involving short or long or mixed GRBs samples, suggesting that the similar processes are responsible for producing the main bursts in two classes of GRBs.
In particular, under the assumption that there is no correlation between the sizes of two events, an exact relation between the index $\alpha$ of the avalanche size distribution and the $q$ value of the appropriate $q$-Gaussian function has been obtained as \citep{Caruso2007,Celikoglu2010,Wei2021}
\begin{equation} \label{eq5}
q=\frac{\alpha+2}{\alpha}.
\end{equation}
Thus, the theoretical $q$ values can be derived as $ q_P = 2.2 $ and $q_T = 2$ by taking the theoretical values of $\alpha$ when $S = 3$. In fact, we find that the $q$ values of precursors are found to be slightly larger than those of main bursts. However, the $q$ values exhibited by both the precursors and the main bursts are smaller than the theoretical ones.

At present, the fact remains that whether precursors are inherently different from main bursts because there is no unambiguous definition of ``precursor''. One is that precursors may indeed be generated from the different physical processes. The other is that precursors possess the similar origins to those of the main bursts and the absence of the dimmer precursors might be
due to the limited sensitivity or energy coverage of the current GRB detectors \citep[e.g.][]{Lxj2021}. In the fireball model, the central engines of GRBs have long-lasting activities and arise the prompt emissions from early internal shocks and X-ray flares from late internal shocks through collisions of relativistic shells ejected. Precursors could be photospheric emission when it is bright enough to
be detected as a separate component in time from the main bursts \cite[e.g.][]{Lyutikov2000}. In this scenario, the precursors are favorable to the
non-thermal emission, which are not well consistent with the previous spectral analysis results \citep{Burlon2008,Hu2014,Lxj2022}. The mean value of quiescent time in the precursor sample is 14.6 s and the longest one is $\sim$ 526 s \citep{Li2022}, which can be explained by multistage collapse model \citep{Lipunova2009} and collapsar fallback model \citep{Wang2007}. In addition, the GRB jet may be magnetically dominated and the magnetic energy can be dissipated by magnetic reconnection based on the quiescent period timescale \citep{Lyutikov2003}. \cite{D2015} reported that GRBs can be produced in a cellular automaton model of a magnetic flux tube with background flow and explained in an SOC framework via magnetic reconnection processes in the magnetized fireball. Moreover, they found that the energetic initial impulses in the magnetohydrodynamic flow
can cause one-dimensional signatures when a two-dimensional SOC regime has been established.
Owing to the radial component of magnetic fields decaying with the radius, the early prompt phases are in a three-dimensional form while the later X-ray flares are closer to one dimension \citep{Wang2013,Lv2020,Lxj6}. Thus, the same Euclidean dimensions of SOC systems between precursor and main burst near to the central engine is reasonable.

Recently, a peculiar long GRB 211211A is found to be associated with a kilonova, implying that the event may originate from a merger of compact star binaries \cite[e.g.][]{Yang2022,Troja2022}. About 1 s prior to the main burst, a precursor flare lasting for $\sim$ 0.2 s presents a feature of quasi-periodic oscillation at frequency $\approx$ 22 Hz, suggesting that the flare results from a resonant shattering of a magnetar before coalescence \citep{Xiao2022,Suvorov2022}. On the other hand, the detection of fast radio burst (FRB) 200428 from the Galactic soft gamma-ray repeater (SGR) J1935+2154 firstly confirms the magnetar origin of some FRBs \citep{CHIME2020,Bochenek2020}. In addition, sub-second periodicity in FRB 20191221A further strengthens the evidence of magnetar origin \citep{Chime2022}. It is worth noting that the power-law size distribution features and scale-invariance structures in FRBs have been reported, indicating that both FRBs and GRBs can be explained by the same physical framework of fractal-diffusive SOC systems with the spatial dimension $S = 3$ \citep{Wang20172,Wei2021,Zhang2021,Wang2023}. Thus, it would be very
intriguing to contain whether the central engines of at least some GRBs are magnetars by the SOC behaviors. Even the physical correlation between GRBs and FRBs can be further studied in the future.

\section{Conclusions} \label{sec:result3}
In this letter, we concentrated on investigating the SOC signatures of precursors in long GRBs. We found that similar scale-free power-law distribution properties and scale-invariance structure of avalanche size differences all exist in precursors, which can be well explained with the SOC framework driven by a magnetically dominated process. In addition, we further inspected the SOC behaviors of main bursts and confirmed that precursors and main bursts have similar statistical properties, indicating that both of them follow likely the same stochastic process and physical mechanism. The similar power-law indexes of size distributions and $q$ values of $q$-Gaussian distributions can be attributed to the same SOC systems with spatial dimension $S = 3$.

\section*{Acknowledgements}
We would like to thank Huang Yong-Feng, Zhang Zhi-Bin and Yi Shuang-Xi for beneficial discussions and helpful
suggestions. We greatly acknowledge the anonymous referees for the valuable comments.  This work is supported by Shandong Provincial Natural Science Foundation (Grant Nos. ZR2023MA049 and ZR2021MA021).

\listofchanges
\end{document}